\documentclass[aps,prb,superscriptaddress,floatfix,twocolumn]{revtex4-2}

\usepackage{bm}        
\usepackage{amsmath,amssymb,amsthm}   
\usepackage{mathtools}
\usepackage{tabularx}
\usepackage{array}
\usepackage{color,soul}
\usepackage{diagbox}
\usepackage{xcolor}
\usepackage{natbib}
\usepackage{siunitx} 
\usepackage{hyperref}
\usepackage{grffile}
\usepackage{graphicx}
\usepackage{float}
\usepackage{chemformula}
\usepackage{gensymb}
\usepackage{multirow}

\definecolor{mygreen}{rgb}{0.1,0.6,.1}
\definecolor{myred}{rgb}{.6,.1,.1}
\definecolor{myblue}{rgb}{.1,.1,.8}
\definecolor{mycyan}{rgb}{0,.8,.8}
\definecolor{mymagenta}{rgb}{0.9,0,0.9}

\newcommand{\wn}{\,cm$^{-1}$}
\newcommand{\Ag}[1]{A$_{\text{g}}^{#1}$}

\newcommand{\Eg}[1]{E$_{\text{g}}^{#1}$}

\newcommand{\TN}{{$T_\textrm{N\'eel}$}}
\newcommand{\Au}[1]{A$_{\text{u}}^{#1}$}
\newcommand{\Eu}[1]{E$_{\text{u}}^{#1}$}

\begin{document}
\title{Spin-Orbital-Lattice Coupling and the Phonon Zeeman Effect in the Dirac Honeycomb Magnet CoTiO$_3$}
\author{Thuc T. Mai}%
\affiliation{Quantum Measurement Division, Physical Measurement Laboratory, NIST, Gaithersburg, MD 20899}
\author{Yufei Li}
\affiliation{Department of Physics, The Ohio State University. Columbus, OH 43210}
\author{K.F. Garrity}%
\affiliation{Materials Measurement Science Division, Materials Measurement Laboratory, NIST, Gaithersburg, MD 20899}
\author{D. Shaw}%
\affiliation{Department of Physics, Colorado State University, Fort Collins, CO 80523}
\author{T. DeLazzer}%
\affiliation{Department of Physics, Colorado State University, Fort Collins, CO 80523}
\author{R.L. Dally}
\affiliation{NIST Center for Neutron Research, National Institute of Standards and Technology, Gaithersburg, MD 20899}
\author{T. Adel}%
\affiliation{Quantum Measurement Division, Physical Measurement Laboratory, NIST, Gaithersburg, MD 20899}
\author{M.F. Mu\~noz}%
\affiliation{Quantum Measurement Division, Physical Measurement Laboratory, NIST, Gaithersburg, MD 20899}
\author{A. Giovannone}%
\affiliation{Department of Physics, The Ohio State University. Columbus, OH 43210}
\author{C. Lyon}%
\affiliation{Department of Physics, The Ohio State University. Columbus, OH 43210}
\author{A. Pawbake}
\affiliation{Laboratoire National des Champs Magnetiques Intenses, LNCMI-EMFL.CNRS UPR3228, Univ. Grenoble Alpes, Univ. Toulouse, Univ. Toulouse 3, INSA-T, Grenoble, France}
\author{C. Faugeras}
\affiliation{Laboratoire National des Champs Magnetiques Intenses, LNCMI-EMFL.CNRS UPR3228, Univ. Grenoble Alpes, Univ. Toulouse, Univ. Toulouse 3, INSA-T, Grenoble, France}
\author{F. Le Mardele}
\affiliation{Laboratoire National des Champs Magnetiques Intenses, LNCMI-EMFL. Grenoble, France}
\author{M. Orlita}
\affiliation{Laboratoire National des Champs Magnetiques Intenses, LNCMI-EMFL. Grenoble, France}
\affiliation{Institute of Physics, Charles University, Ke Karlovu 5, 121 16 Prague, Czech Republic}
\author{J.R. Simpson}%
\affiliation{Department of Physics, Astronomy, and Geosciences, Towson University, Towson, MD 21252}
\affiliation{Quantum Measurement Division, Physical Measurement Laboratory, NIST, Gaithersburg, MD 20899}
\author{K. Ross}%
\affiliation{Department of Physics, Colorado State University, Fort Collins, CO 80523}
\author{R. Vald\'es Aguilar}
\email{rvaldesag@gmail.com}
\affiliation{Department of Physics, The Ohio State University. Columbus, OH 43210}
\author{A.R. Hight Walker}
\email{angela.hightwalker@nist.gov}
\affiliation{Quantum Measurement Division, Physical Measurement Laboratory, NIST, Gaithersburg, MD 20899}
\date{\today}
\begin{abstract}
The entanglement of electronic spin and orbital degrees of freedom is often the precursor to emergent behaviors in condensed matter systems. With considerable spin-orbit coupling strength, the cobalt atom on a honeycomb lattice offers a platform that can make accessible the study of novel magnetic ground states. Using temperature-dependent Raman spectroscopy and high-magnetic field Raman and infrared (IR) spectroscopy, we studied the lattice and spin-orbital excitations in \ch{CoTiO_3}, an antiferromagnetic material that exhibits topologically protected magnon Dirac crossings in the Brillouin zone. Under the application of an external magnetic field up to 22 T along the crystal's $c$-axis, we observed the splitting of both the spin-orbital excitations and a phonon nearby in energy. Using density functional theory (DFT), we identify a number of new modes that below the antiferromagnetic (AFM) transition become Raman-active due to the zone-folding of the Brillouin zone caused by the doubling of the magnetic unit cell. We use a model that includes both the spin and orbital degrees of freedom of the Co$^{2+}$ ions to explain the spin-orbital excitation energies and their behavior in an applied field. Our experimental observations along with several deviations from the model behavior point to significant coupling between the spin-orbital and the lattice excitations.
\end{abstract}
\maketitle

\section{Introduction}
\begin{figure}[t]
\includegraphics[width=0.99\columnwidth]{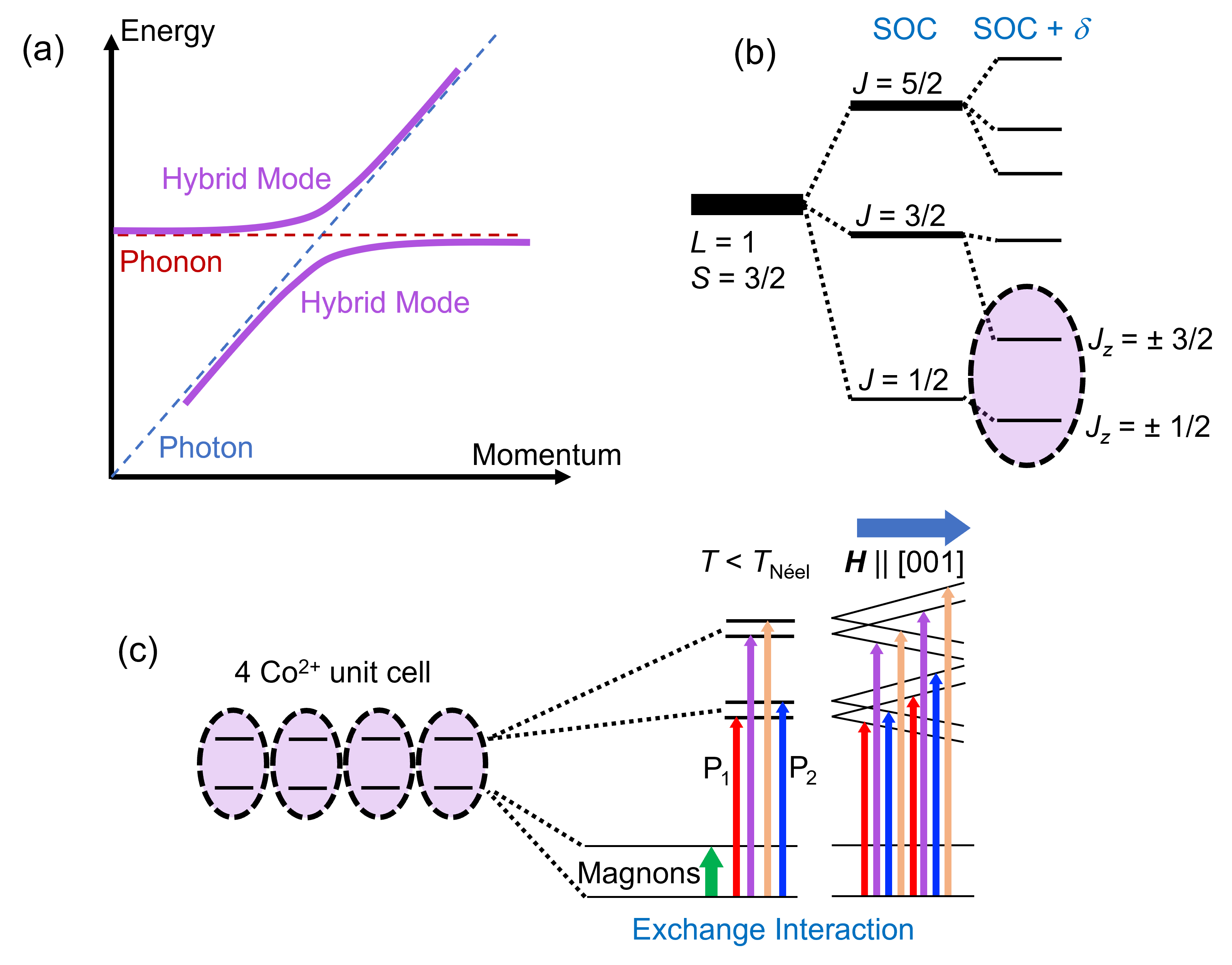}
\caption{\textbf{Hybrid Modes and Spin Orbitons in CoTiO$_3$}.(a) Schematic of an avoided level crossing between phonon and photon interacting to generate hybridized modes. (b) Energy level diagram (not to scale) of a single Co$^{2+}$ ion with $S=3/2$ and effective orbital angular momentum $l=1$. The multiplet is split by spin-orbit coupling and the trigonal distortion of the oxygen octahedron surrounding the Co atom. The two lowest doublets are highlighted in lavender colored oval. (c) Combination of the two lowest doublets from 4 Co atoms (two on each honeycomb layers) to generate new energy levels in the magnetically ordered state. The new first excited state corresponds to the typical magnon modes that result from the exchange interactions considered in ref. \cite{Yufei}. The next two excited states result from the combination of the 4 pairs of $J_z=\pm3/2$ states. These can then split under the application of a magnetic field along the $c$-axis. Red/blue (purple/orange) arrows mark the Raman (infrared) transitions possible between the ground state and the excited states.}
\label{fig:1}
\end{figure}

Quasiparticles, which encompass dressed elementary particles (i.e. electrons), composite objects (Cooper pairs), and collective excitations (phonons, magnons), are of fundamental interest in condensed matter physics. Energy, linear momentum, and their unique inter-relationship, the so-called dispersion relation, characterize such quasiparticles. When the symmetry of a crystal allows it, quasiparticles can interact with each other and create hybrid modes whose wave functions inherit properties of the parent quasiparticles. An example of this hybridization is the interaction between photons and polar phonons in insulators \cite{PhysRevLett.15.964,PolaritonPhysics,TerahertzPolaritonics,Mai2021_SciAdv}. This hybridization can be seen most clearly near avoided crossings of the     hybridized modes as a parameter of the system like momentum, magnetic field, or temperature, is varied. This is schematically shown in Fig. \ref{fig:1}(a), where the bare, non-interacting quasiparticles' dispersions are shown with dashed lines, and the hybridized modes with solid lines.

Magnetically ordered materials offer a fertile ground for the observation of hybridization between quasiparticles that such systems can host. In insulating materials, magnetic order typically occurs because of the interaction between pure atomic spins with their neighbors. However, it is also possible that the atomic magnetic degrees of freedom that are not of pure spin origin, such as orbital angular momentum, play a role in the magnetic behavior as well. This interplay originates from the electric field generated by the atoms surrounding the magnetic atom, usually called crystal electric field (CEF), and its effect on the electronic energy levels of the atoms in the unit cell of the material \cite{AandBbook,tanabesugano, Khomskii_2014}. The combination of the CEF, spin-orbit coupling (SOC), small structural distortions, and the interactions between atoms can produce a series of energy levels that must be considered to explain all of the magnetic properties of a material. A recent example of this behavior is \ch{CoTiO_3}, where both magnons and these higher energy spin-orbital states have been found to have dispersion in momentum space \cite{YuanPRX2020,Elliot2021}, and are required to explain many magnetic properties \cite{Yufei}. These spin-orbital states are sometimes called spin-orbit excitons\cite{YuanPRB2020,Elliot2021} or spin-orbitons\cite{spinOrbiton_2015} because they are true quasiparticles with specific energy versus momentum dispersion relations. We use the spin-orbiton nomenclature in this article. 

For third row transition metal oxides, the energy scales of SOC and lattice distortions \cite{Khomskii_2014} can make the spin-orbitons energies overlap with optical phonons (10 meV to 100 meV). The coupling between these quasiparticles can then lead to the so-called phonon Zeeman effect \cite{Juraschek_2019}, the splitting of degenerate phonons with an applied magnetic field. Several microscopic models predict this effect in large classes of materials \cite{Juraschek2017,Juraschek_2019,Juraschek2020}, but it has been unclear which one dominates in specific materials. At the same time, this effect has been observed in CeF$_3$, NdF$_3$, CeCl$_3$ \cite{schaack_magnetic-field_1975,Schaack_1976,Schaack1977} where doubly degenerate phonons split in a magnetic field due to their coupling to transitions between the CEF-split 4f electronic levels of the rare earth atom. 

From a group theoretical perspective, we can understand the phonon Zeeman effect as coming from the effect that time-reversal symmetry breaking has on the phonons. In a crystalline material, all elementary quasiparticles can be classified according to the irreducible representation of its point group. However, when the irreducible representations have complex characters\cite{ITCD1,charactertable}, it is a common practice in physics \cite{ITCD2} to combine complex conjugated representations into so-called physical irreducible representations with real characters. These combined physical representations will be doubly degenerate, and are used to label quasiparticles at the Brillouin zone center whose degeneracy is either accidental or protected by time-reversal symmetry. This occurs in 10 crystallographic point groups that host doubly degenerate reducible representations with complex characters \cite{Ovander1960}. These point groups are: $4$, $\bar{4}$, $4/$m, $3$, $\bar{3}$, $\bar{6}$, 6, 6/m, $23$, m$\bar{3}$. Two of them are cubic (23, m$\bar{3}$), and the other eight are uniaxial \cite{ANASTASSAKIS1972}. If time-reversal symmetry is not broken, there will be pairs of degenerate zone-center phonons with complex characters that are conjugates of each other, and are labeled with letter 'E', e.g., in point group $\bar{3}$ they are labeled \Eg{} and \Eu{} for inversion-even and odd modes, respectively. The degeneracy between the pair is enforced by time-reversal symmetry. Thus, an applied magnetic field removes the degeneracy and splits the phonon energies. Since \ch{CoTiO_3} belongs to the uniaxial crystal class $\bar{3}$, we expect that some of its \Eg{} phonons will split under the application of a magnetic field.

\ch{CoTiO_3} has the ilmenite structure, with space group of R$\bar{3}$\cite{Newnham:a04103,YuanPRX2020,Elliot2021}. Using the hexagonal notation, in the $a$-$b$ plane, the Co$^{2+}$ ions are arranged in honeycomb lattices with edge-sharing oxygen octahedra. Interestingly, such d$^7$ electronic systems have been proposed as promising candidates for the realization of the Kitaev model with its quantum spin liquid ground state \cite{liuKhaliullin}. However, below \TN = 38\,K, the Co$^{2+}$ spins in \ch{CoTiO_3} magnetically order, with ferromagnetic spins within each honeycomb plane antiferromagnetically aligned along the $c$-axis. The spin direction is parallel to the $a$-$b$ plane. According to previous inelastic neutron studies, the ground state with an effective spin $J_{\text{eff}}=1/2$ describes qualitatively well the magnon spectrum \cite{YuanPRX2020,Elliot2021} except for the magnon gap at the zone center. This gap can be explained by a ring-exchange interaction between the spins in the honeycomb plane \cite{Yufei}. The neutron studies also found excitations from the ground state into higher energy $J_{\text{eff}}$ states. A flavor wave model describes both spin and orbital contributions to the magnetic excitation spectrum \cite{Elliot2021,Yufei}.

\begin{figure*}[]
\includegraphics[width=1\textwidth]{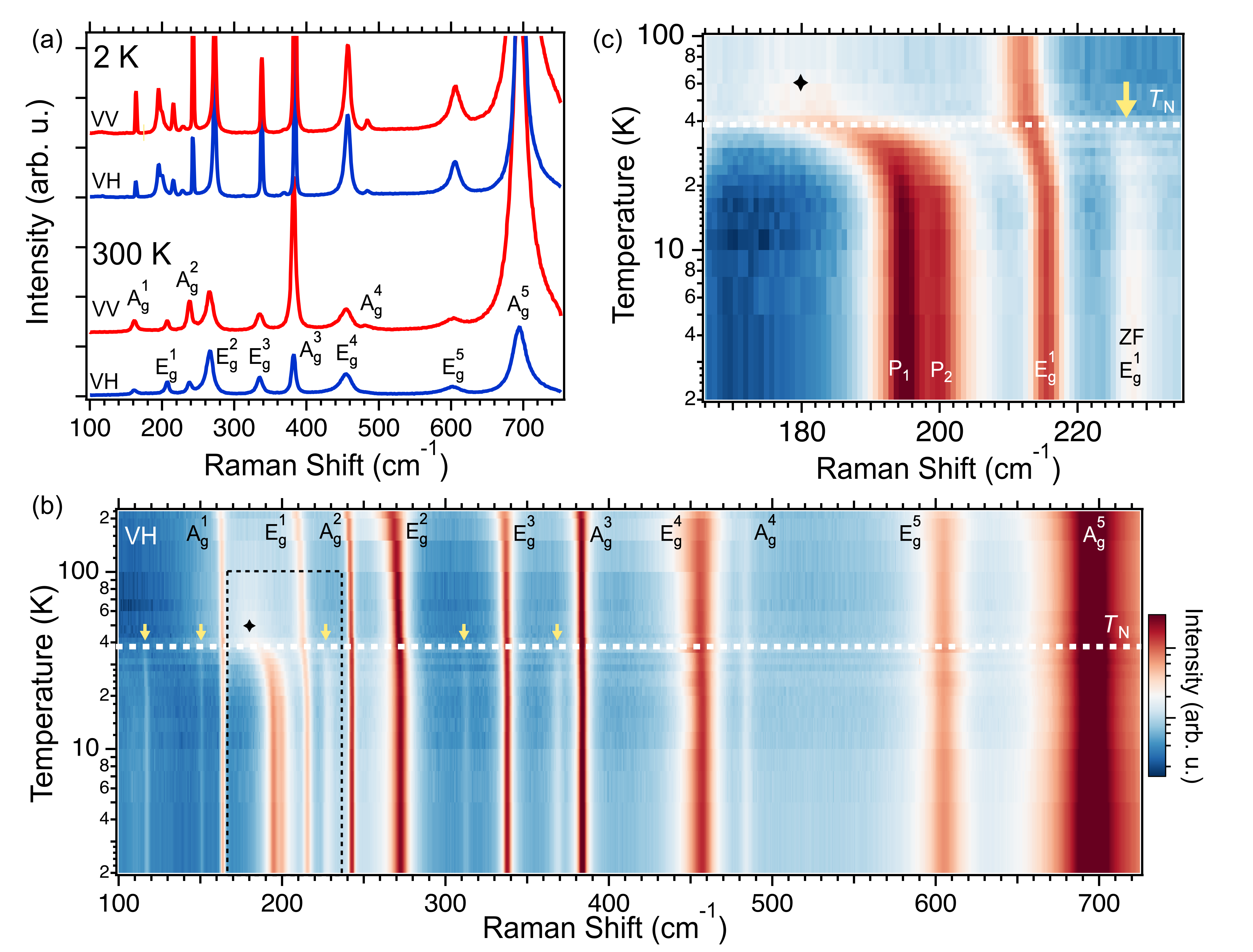}
\caption{\textbf{Raman spectra of CoTiO$_3$}. (a) The Raman spectra for parallel (VV) and crossed (VH) polarizations collected with 488\,nm excitation laser at 300\,K (bottom) and 2\,K (top). Phonon modes are labeled with representations of the point group $\overline{3}$ at room temperature. (b) False color intensity plot of the temperature dependent Raman spectra in VH configuration. The dashed horizontal line represents the N\'eel temperature of 38\,K. Arrows highlight the emergence of new modes below \TN. The diamond marks the high temperature energy of the lowest spin-orbital transition. The interaction between the spin-orbitons P$_1$ and P$_2$ and nearby \Eg{1} phonon as well as its zone-folded (ZF) counterpart is highlighted in the dashed rectangle and expanded in panel (c).}
\label{fig:Tdep}
\end{figure*}

Within a single Co$^{2+}$ ion, in a trigonally distorted octahedral environment, there are three main energy scales: crystal field splitting ($E_{\text{cf}}$), trigonal distortion ($\delta$), and SOC ($\lambda$), where $E_{\text{cf}} \gg \delta\gtrsim\lambda$ \cite{YuanPRB2020,Elliot2021}. The 3d$^7$ electrons from a free Co$^{2+}$ ion have their degeneracy lifted by the three aforementioned effects. As a result, the lowest lying energy states are six doubly degenerate spin-orbital levels, as shown in Fig. \ref{fig:1}(b), that can be labeled by the total angular momentum $J_z$ quantum number, where $z$ is picked out by the direction of the trigonal distortion, and is the same as the crystallographic $c$-axis. Above \TN\, the ground state to the first excited state transition, $J_z=\pm1/2$ to $J_z=\pm3/2$, is approximately 23 meV, or 185\wn \cite{YuanPRB2020,Elliot2021}. This transition is also observed in the Raman spectrum as described in this article.

Below the magnetically ordering temperature, the magnetic interactions between the four Co atoms in the unit cell will produce a new energy level structure, as schematically shown in Fig. \ref{fig:1}(c). The doublets with $J_z=\pm3/2$ will combine to produce eight new levels whose energies will depend on all of the magnetic interactions, such as those considered in ref. \cite{Yufei}. Some of these levels have been found to have dispersion in momentum space \cite{YuanPRX2020,Elliot2021}, making them true quasiparticles of the magnetic system, labeled as spin-orbitons. Their energies can be modified by an applied magnetic field. Recently, the coupling between spin-orbitons and doubly degenerate phonons of E$_\text{g}$ symmetry in CoTiO$_3$ was explored \cite{Chaudhary_PRB2024}, and small splittings of the phonon frequencies were predicted.

In this study, we provide clear evidence of the hybridization of spin-orbitons and a zone-center, doubly degenerate E$_\text{g}$ phonon via the phonon Zeeman effect measured using magneto-Raman scattering. On the other hand, the infrared active, zone-center phonons did not show any splitting in magnetic field. We also find several phonons that result from the Brillouin zone folding in the magnetically ordered state. Surprisingly, a zone-folded phonon also exhibit the behavior of the phonon-Zeeman effect, thus highlighting the significant coupling between the lattice and the spin-orbitons.

\begin{figure*}[]
\includegraphics[width=1.9\columnwidth]{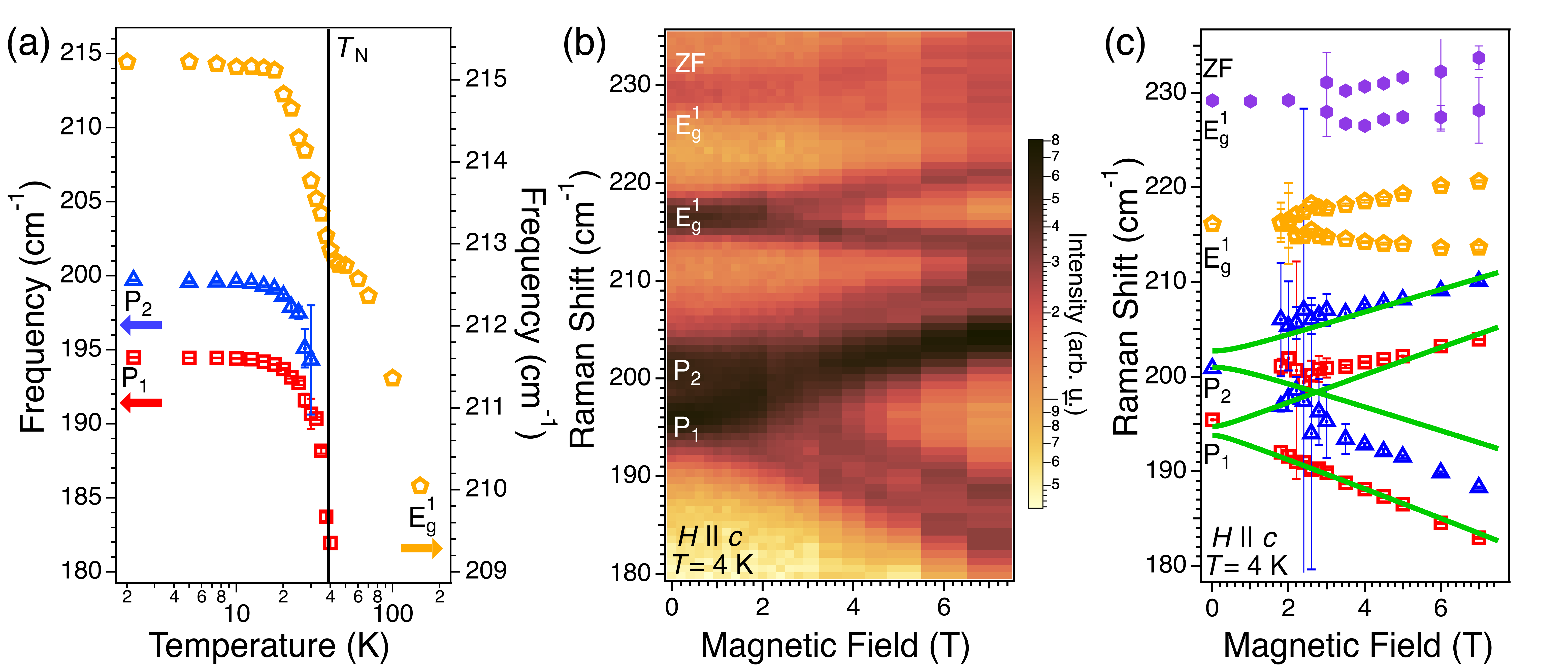}
\caption{\textbf{Interaction between spin orbitons P$_1$ and P$_2$ with E$_\text{g}^1$ phonon: Phonon Zeeman Effect}. (a) Temperature dependence of frequencies of spin orbitons and E$_\text{g}^{1}$ phonon at zero magnetic field. \TN { }is marked by the dark vertical line. The error bar corresponds to 1 standard deviation obtained from the fitting. (b) False color map of Raman scattering intensity as a function of magnetic field parallel to the $c$-axis up to 7 T at 4 K, including the zone-folded E$_\text{g}^1$ phonon. The modes P$_1$ and P$_2$, of spin-orbital origin, each splits into two. The \Eg{1} phonon and its zone-folded relative can be seen splitting and broadened, respectively. (c) Field dependence of the frequencies at 4 K of the spin-orbitons P$_1$ and P$_2$, and of the phonon E$_\text{g}^1$. Solid lines are calculations of the spin-orbitons frequencies using the model of \cite{Yufei}.}
\label{fig:Fitting}
\end{figure*}

\section{Results}

We observed 5E$_{\text{g}}$ + 5A$_{\text{g}}$ Raman active phonons in the $c$-cut crystal at room temperature as shown at the bottom of Fig. \ref{fig:Tdep}(a), in both VV and VH polarization configurations. This is as expected from group theoretical analysis and consistent with a previous experiment on \ch{CoTiO_3} \cite{DUBROVIN2021}. As the sample is cooled down, the Raman active phonon frequencies increase as expected from the anharmonic contraction of the lattice, and below \TN\ the spectrum is richer as shown at the top of Fig. \ref{fig:Tdep}(a). The entire temperature dependence of the Raman phonon spectrum is shown as a false color intensity map in Fig. \ref{fig:Tdep}(b). At \TN, there are several phonons that appear in the spectrum as a result from the folding of the Brillouin zone due to the doubling of the unit cell in the antiferromagnetic state, as we will show below. These are identified with arrows in Fig. \ref{fig:Tdep}(b). 

One feature that becomes measurable as a broad mode centered around $\approx\qty{180}{\centi\meter^{-1}}$ below 100\,K (identified by a diamond in Fig. \ref{fig:Tdep}(b)). The energy of this mode matches the transition energy between the spin-orbital ground state and the first excited doublet previously measured \cite{YuanPRB2020,YuanPRX2020,Elliot2021}. This mode has a strong change in energy below the magnetic transition temperature, \TN; it splits into at least two modes and their frequencies reach $\approx$ 195\wn and 201\wn\ at 2 K, indicating a strong renormalization due to the magnetic order. This is consistent with the picture of the creation of new levels due to the interaction of the excited doublets from the four Co atoms in the magnetic unit cell, as schematically shown in Fig. \ref{fig:1}(c). In the magnetically ordered state, the exchange interactions renormalize the energies of these levels, just like they do for the magnons at the zone center \cite{Yufei}. We label these two modes as P$_1$ and P$_2$ and identify them with the spin-orbitons measured with inelastic neutron scattering \cite{YuanPRX2020,Elliot2021}. We show this behavior in Fig. \ref{fig:Tdep}(c). Due to the higher energy resolution in our Raman measurement, we are able to resolve the finer structure at the Brillouin zone center, $\bm{k}=0$, of the spin-orbitons.

We fit the temperature dependence of the frequencies of P$_1$ and P$_2$ to better understand their behavior across the magnetic transition. The results are shown in Fig. \ref{fig:Fitting}(a) together with the frequency of the nearby \Eg{1} phonon. The magnitude of the change in frequency in P$_1$ and P$_2$ (Fig. \ref{fig:Fitting}(a)) is comparable to a magnon's as reported in \cite{Yufei}. As the temperature increases towards \TN, both P$_1$ and P$_2$ experience a shift to a lower frequency of more than 8\wn (1 meV). For comparison, the \Ag{1} phonon shifted less than $\qty{1}{\centi\meter^{-1}}$ over our measurement temperature range (not shown). Above 30\,K, both P$_1$ and P$_2$ become too broad and hence impossible for the fitting procedure to separately resolve. The correlation between the dramatic changes in P$_1$ and P$_2$ with the magnon temperature dependence implies that these modes are renormalized due to the same exchange interactions that determine the behavior of the magnons. While this large shift in the phonon frequency of \Eg{1}, with its clear change at \TN , could be interpreted as this phonon being coupled to the magnetic order parameter (spin-phonon coupling), it is better understood as originating from the strong coupling and hybridization with the nearby spin-orbiton modes, P$_1$ and P$_2$, a particular type of spin-orbital-lattice coupling. We make this inference because none of the other E$_\text{g}$ or A$_\text{g}$ phonons show this behavior and the lattice anharmonicity is already frozen out at these temperatures and thus no phonon frequency changes would be expected.

\begin{figure}[t]
\includegraphics[width=1\columnwidth]{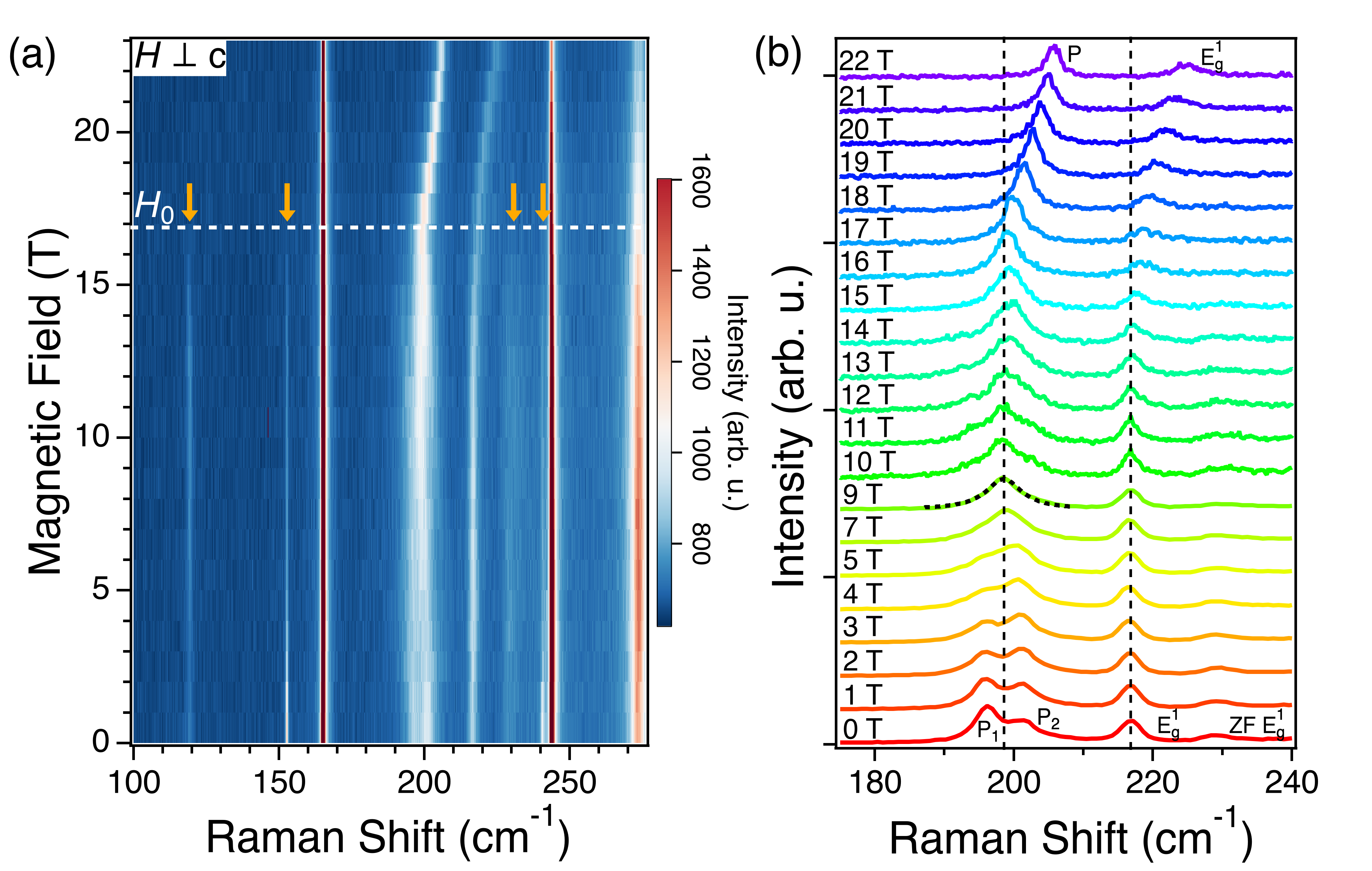}
\caption{\textbf{High Magnetic Field Behavior}. (a) False color Raman scattering intensity map at 5 K as a function of magnetic field applied perpendicular to the $c$-axis. Zone-folded phonons marked by arrows disappear above the critical field $H_0$. (b) The same data as in (a), but now focused in the region of P$_1$, P$_2$, and E$_\text{g}^1$. Data for 9 T or lower were collected at the National Institute of Standards and Technology (NIST) with 488 nm, and above 9 T at the Laboratoire National des Champs Magn\'etique Intenses (LNCMI) in Grenoble, France with 633 nm. Black dashed line at 9 T shows the fit to a single mode P with Lorentzian line shape, the merged P$_1$ and P$_2$ spin-orbiton. Above $H_0$, both P and E$_\text{g}^1$ shift with magnetic field. Dashed vertical lines mark the positions of P and E$_\text{g}^1$ at 9 T.}
\label{fig:FieldDep}
\end{figure}

We now focus on the spectral region containing P$_1$, P$_2$, and the \Eg{1} phonon for out of plane magnetic field, $\textit{\textbf{H}} \parallel c$, shown in Fig. \ref{fig:Fitting}(b) and (c). Na\"ively, we expect the spin-orbitons with finite Lande $g$-factor\cite{Elliot2021} to couple to a Zeeman field. Suprisingly, we observe the splitting of both P$_1$ and P$_2$, as well as the \Eg{1} phonon (Fig. \ref{fig:Fitting}(b)). There are no noticeable changes in the rest of the phonon spectrum with the exception of the zone-folded version of \Eg{1}, around 230\wn, which appears to also split or significantly broaden with increasing magnetic field magnitude. These splittings are the key signature for identifying the phonon Zeeman effect \cite{Juraschek_2019} and are one of the characteristics of chiral phonons. The same model used in ref.\cite{Yufei} is able to reproduce the magnetic field splitting of the spin-orbitons (solid lines in Fig. \ref{fig:Fitting}(c)), but it does not consider the nearby phonons.  While the splitting of a phonon is highly unusual, we take it as the clearest signature of the hybridization between the spin-orbitons and the \Eg{1} phonon, confirming our previous inference.

Next, we study the spectrum in a magnetic field perpendicular to the honeycomb plane in the magnetically ordered state (5 K). Fig. \ref{fig:FieldDep}(a) shows the data with magnetic field parallel to the [110] direction up to 22 T showing four identifiable modes that disappear from the spectrum above a field $H_0\approx 17$ T. These modes are the same modes that appear below \TN\ in the temperature dependent data in Fig. \ref{fig:Tdep}(c). The transition at $H_0$ is between the antiferromagnetic state and a fully polarized paramagnet with moments parallel to the external magnetic field \cite{permeability2021}. This transition reduces the size of the magnetic unit cell back to the crystallographic one, containing only two Co spins and all the phonons that become active in the antiferromagnetic state with a doubled unit cell have to disappear from the zone center. This confirms that these extra modes are zone-folded phonons.

With this direction of the external magnetic field, we observe no splitting on any of the phonons or spin-orbitons. Surprisingly, we measured what appears to be the merging of P$_1$ and P$_2$ around 7 T (Fig. \ref{fig:FieldDep}(b)). At 9 T and up to $H_0$, we can no longer separately fit P$_1$ and P$_2$, as seen by the fit of a single peak line shape (black dashed line). This behavior continues for the newly merged P spin-orbiton up to $H_0$, above which the behavior changes to a linear shift of its frequency. This shift also occurs in the nearby \Eg{1} phonon. The \Eg{1} phonon frequency does not shift for fields below $H_0$, and the fact that this shift occurs above $H_0$ provide further evidence for the hybridization of this \Eg{1} and the spin-orbiton; as soon as the P spin-orbiton shifts, the \Eg{1} phonon responds by \textit{moving away} from the spin-orbiton. 

\begin{figure}[b]
\includegraphics[width=1.0\columnwidth]{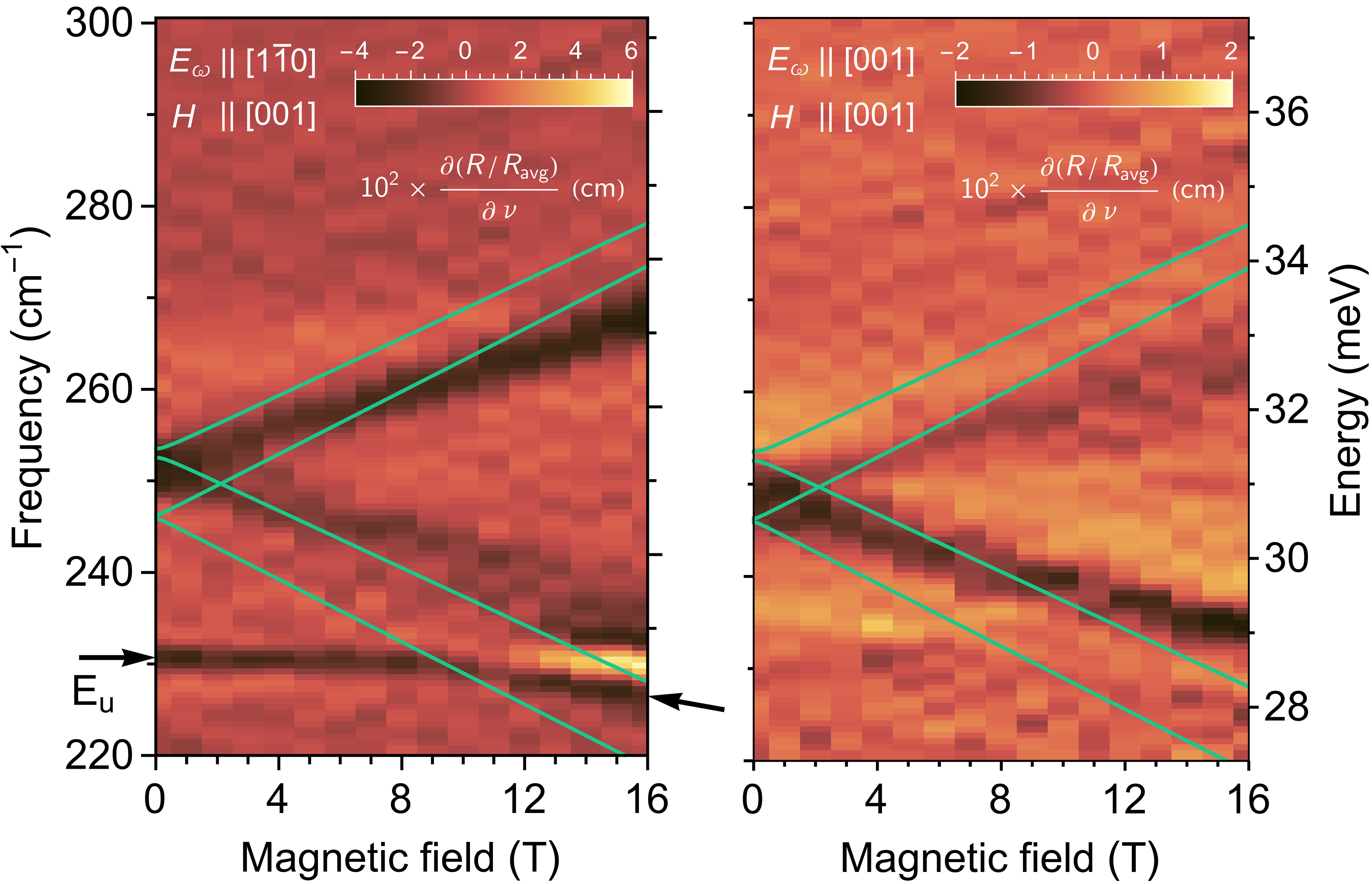}
\caption{\textbf{High Magnetic Field Far Infrared Reflection}. False color map from infrared reflection derivative. In both plots, the raw reflectivity is divided by the average of all data at different fields. Then the partial derivative $\frac{\partial\left(R_B/R_{\text{avg}}\right)}{\partial \nu}$ is plotted. In both experiments, the external static magnetic field is applied along the $c$-axis. The polarizations in these experiments are aligned parallel to $\left[1\overline{1}0\right]$ and $[001]$ in panel (a) and (b), respectively. Green lines are from results from the model in \cite{Yufei}. The arrows in (a) indicate an \Eu{} phonon.}
\label{fig:FTIRFieldDep}
\end{figure}

Finally, we explore the possibility of hybridization between spin-orbitons and phonons with infrared spectroscopy. In this experiment, only excitations which are electric-dipole active (i.e. inversion odd) can be detected. Phonons of this nature in \ch{CoTiO_3} will have labels of \Eu{} and \Au{} symmetries. However, because only \Eu{} phonons appear in complex conjugate pairs in the point group of \ch{CoTiO_3}, we do not consider the \Au{} phonons. We expect a similar effect on the \Eu{} phonons as was observed in the \Eg{} ones with Raman spectroscopy. In Fig. \ref{fig:FTIRFieldDep}, we show the reflection derivative with respect to frequency $\frac{\partial\left(R_B/R_{\text{avg}}\right)}{\partial \nu}$ obtained at 4.2\,K and normalized to the reflection averaged over all magnetic fields. We focus on the energy range near an \Eu{} phonon around 231\wn \cite{DUBROVIN2021} and a nearby infrared active spin-orbiton around 251\wn \cite{YuanPRB2020,Elliot2021}. This high energy spin orbiton corresponds to the highest energy quadruple level shown in Fig. \ref{fig:1}(c). Because we could not measure a proper reference to obtain the absolute value of the reflectivity, we choose to show the derivative of the measured reflection because its minimum is close to the frequency of transverse excitations when they are weak and narrow. This allows us to track the field dependence of the excitations without using a model. This is shown in both panels of Fig. \ref{fig:FTIRFieldDep} for the two measured polarizations, (a) for $E_\omega||[1\bar{1}0]$ and (b) for $E_\omega||[001]$. In both cases the static magnetic field is parallel to the $c$-axis.

The data show that the spin-orbiton at 251\wn splits into two modes as shown by the minimum (dark color) of the reflection derivative in both polarization configurations. In these same plots we have added the prediction from the previously used theory \cite{Yufei} as green lines. The zero field value of the spin-orbiton is well reproduced by the theory, but the field dependence is less so. The model, briefly described in the next section, predicts a significant zero field splitting of this spin-orbiton and it also predicts larger slopes (effective g-factors) for these four excitations. Interestingly, we observe the \Eu{} phonon (only observed in panel (a)) shifts with the application of the magnetic field, and in particular it appears that the lower branch of the spin orbiton and the phonon undergo an avoided crossing as the phonon shifts down more than 4\wn\ above $\approx$ 9\,T. This is further indication of hybridization between the \Eu{} phonon and this high energy spin-orbiton, but it is not consistent with the phonon being a chiral phonon as it does not split in magnetic field.
In the next section, we will discuss the implications of these observations and their possible explanations.

\begin{figure}[b]
\includegraphics[width=.48\textwidth]{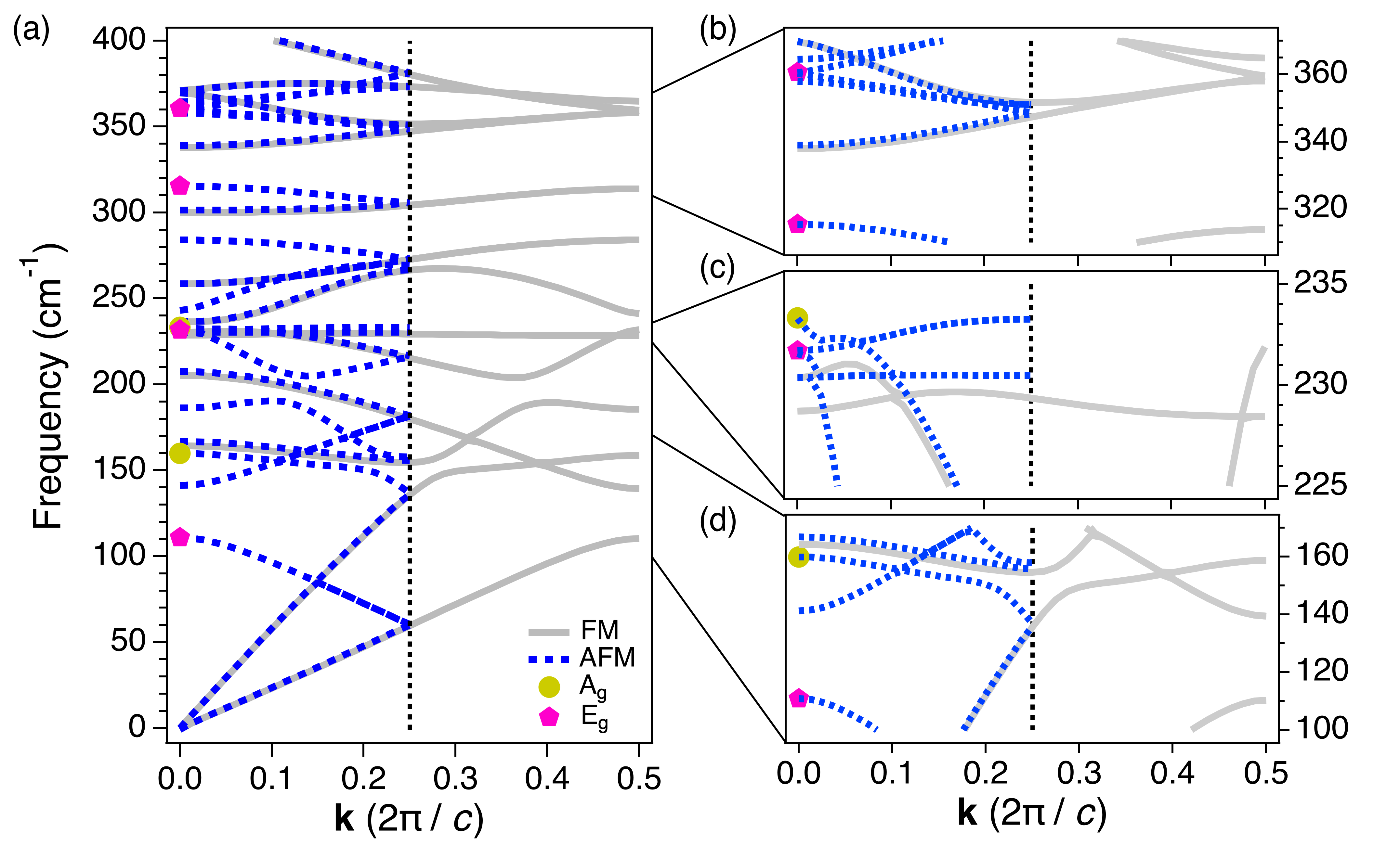}
\caption{\textbf{Calculated Raman phonon dispersion in CoTiO$_3$}. (a) Wide frequency range phonon dispersion along the $c$-axis for AFM and FM states. Zone-folded phonon frequencies observed in the Raman experiments are marked by symbols (circles and pentagons for A$_\text{g}$ and E$_\text{g}$ symmetries, respectively). (b), (c), and (d) show details of the phonon dispersion near the frequencies of the measured zone-folded phonons.
}
\label{fig:dft-phonons}
\end{figure}

\section{Discussion}

We have described three main results in the high energy Raman and infrared spectra in \ch{CoTiO_3}: 1) the appearance as a function of temperature, and disappearance as a function of in-plane magnetic field above $H_0$ of several zone-folded phonons; 2) the hybridization between the spin-orbitons P$_1$ and P$_2$ with nearby \Eg{1} phonon demonstrated with the changes in the temperature dependence of their energies at \TN\,, their Zeeman splitting with $H||c$-axis, and a linear shift of P$_1$ and \Eg{1} phonon with $H\perp c$ above $H_0$; and 3) an  additional signature of hybridization between a \Eu{} phonon and an infrared active spin-orbiton in the reflection spectra as a function of magnetic field applied along the $c$-axis. Below, we provide further justification for our conclusions and discuss their limitations.

\renewcommand{\arraystretch}{1.2}
\begin{table}[b]
\caption{\textbf{Raman phonon frequencies}. The experimental phonon peak frequencies at 2\,K from fitting the spectra using a Voigt line shape compared with calculated values from DFT for the Raman active phonons only.}
\begin{tabular}{|W{c}{45pt}|W{c}{70pt}|W{c}{70pt}|}
\hline
Symmetry      &  DFT (\wn)   &   Exp. (\wn) \\ \hline
A$_{\text{g}}^1$       &  166.8		  &   163.4	\\ \hline
E$_{\text{g}}^1$		  &	 207.5		  &	  215.2	\\ \hline
A$_{\text{g}}^2$		  &	 236.6		  &	  242.4	\\ \hline
E$_{\text{g}}^2$		  &	 258.6		  &	  272.1	\\ \hline
E$_{\text{g}}^3$		  &	 339.0		  &	  337.5	\\ \hline
A$_{\text{g}}^3$		  &	 370.8		  &	  383.5	\\ \hline
E$_{\text{g}}^4$		  &	 442.6		  &	  456.8	\\ \hline
A$_{\text{g}}^4$		  &	 468.2		  &	  483.8	\\ \hline
E$_{\text{g}}^5$		  &	 581.3		  &	  604.8	\\ \hline
\Ag{5}		  &	 709.7		  &	  695.3	\\ \hline
\end{tabular}
\label{tab:RamanModes}
\end{table}

\begin{table}[th]
\caption{\textbf{Zone-folded Raman phonon frequencies}. The experimental results at 2\,K compared with calculated values from DFT. $^\dagger$This mode was only observed with unpolarized light. x marks unobserved modes.}
\begin{tabular}{|W{c}{45pt}|W{c}{70pt}|W{c}{70pt}|}
\hline
Symmetry      &  DFT (\wn)   &   Exp. (\wn)\\ \hline
E$_{\text{g}}$        &  110.9		  &   117	\\ \hline
A$_{\text{g}}$		  &	 159.9		  &	  152	\\ \hline
E$_{\text{g}}$		  &	 231.7		  &	  228	\\ \hline
A$_{\text{g}}$		  &	 233.3		  &	  241$^\dagger$     \\ \hline
E$_{\text{g}}$		  &	 315.4        &	  312	\\ \hline
A$_{\text{g}}$		  &	 357.9		  &	  x   	\\ \hline
E$_{\text{g}}$		  &	 360.6		  &	  369	\\ \hline
E$_{\text{g}}$		  &	 537.8		  &	  x 	\\ \hline
A$_{\text{g}}$		  &	 590.2		  &	  x 	\\ \hline
A$_{\text{g}}$		  &	 655.2		  &	  x 	\\ \hline
\end{tabular}
\label{tab:foldedRamanModes}
\end{table}
\renewcommand{\arraystretch}{1}

\subsection{Zone Folding}

To justify the conclusion of the appearance of zone-folded phonons, we performed first principles density functional theory (DFT)\cite{hk,ks,phonopyNew} calculations using the Quantum Espresso code\cite{qe}, with GBRV pseudopotentials\cite{gbrv}. We use the PBEsol exchange correlation function functional\cite{pbesol} and a Hubbard U correction of 3 eV on the Co-d states (DFT+U)\cite{DFTU}. We use the phonopy code to calculate phonon frequencies and perform symmetry analysis \cite{phonopy-phono3py-JPCM,phonopy-phono3py-JPSJ}. We validated these calculations with the low temperature values of the phonon frequencies for the phonons that are present above \TN. The calculated and experimental phonon frequencies in Table \ref{tab:RamanModes} are in excellent agreement, giving confidence to the reliability of the calculations.

We estimate the frequency of potential zone-folded phonons by calculating their dispersion along the $c$-axis, the direction of the unit cell doubling in the AFM state. We calculate the phonon dispersion in a putative FM state in order to keep the unit cell with only two Co atoms, and then calculate it in the AFM state with four Co atoms. The dispersion along $k_z$ of the phonons below 400 \wn\ is shown in Fig.\ref{fig:dft-phonons}(a) with solid lines for the FM and dashed ones for the AFM state. The frequencies of the AFM phonons that match the observed new modes are marked with circles and pentagons at $k_z=0$ for A$_\text{g}$ and E$_\text{g}$ symmetries, respectively. These symmetries are identified using phonopy \cite{phonopy-phono3py-JPCM,phonopy-phono3py-JPSJ}. We expect five new E$_\text{g}$ and four new A$_\text{g}$ phonons with frequencies listed in Table \ref{tab:foldedRamanModes}. We observed four new E$_\text{g}$ and one A$_\text{g}$ phonons and their frequencies match extremely well the calculated values. It remains to be tested whether the other zone-folded phonons with infrared activity (not listed here) are observable too.

We expand the details of the dispersion near the frequencies of the zone-folded Raman phonons in Fig.\ref{fig:dft-phonons}(b), (c), and (d). For the high and low frequency phonons (panels (b) and (d)) the simple zone-folding picture is consistent with the calculations, i.e., the AFM dispersion is precisely the same as in the FM near the edge of the Brillouin zone. On the other hand, panel (c) shows a striking difference between the AFM and the zone-folded FM dispersion. The AFM phonons have a slightly larger frequency than predicted from simply zone-folding of the FM phonons. It is interesting that this occurs for the phonons that appear to be coupled to the spin-orbitons, indicating that the DFT calculation is able to capture some aspect of this coupling from first principles. This confirms that all the extra modes observed below \TN\ and that disappear above $H_0$ are zone-folded phonons and not any other type of excitation.

\subsection{Hybridization between Phonons and Spin Orbitons}
Within the single ion picture, we only expect to see a single, 4-fold degenerate transition between $J_z =\pm 1/2$ and $J_z = \pm 3/2$. A magnetic field applied along the quantization axis, the $c$-axis, will be able to lift the Kramer's degeneracy in these doublets, generating four different excitations. It is evident from the existence of two well separated peaks, P$_1$ and P$_2$, and an additional infrared active one at zero magnetic field that the single ion picture is not accurate at low temperature. In fact, from the measured bandwidth of the dispersion of these spin-orbitons \cite{YuanPRB2020,Elliot2021}, we already know that the magnetic interactions between the Co atoms will significantly modify the single-ion picture.

To capture the splitting between P$_1$ and P$_2$, the energy of the infrared active mode, and their magnetic field dependence, we use a magnetic Hamiltonian that takes into account both the spin and orbital degree of freedom of Co$^{2+}$ ions in \ch{CoTiO_3} lattice \cite{Elliot2021,Yufei}. This flavor wave model was used previously to successfully explain the behavior of magnons, including the origin of the zone-center gap and the magnon magnetic field dependence \cite{Yufei}. In this model, each Co$^{2+}$ is represented as a vector in a 12-dimensional space, spanned by six doublets, with total spin $S=3/2$ and effective orbital angular momentum $l=1$, as shown in Fig. \ref{fig:1}(b). The dimensionality of this single-ion Hilbert space comes from the product of the spin and orbital angular momentum degeneracies, $(2S+1)\times(2l+1)$. Each ion has spin-orbit coupling $(3/2)\lambda \bm{l}\cdot\bm{S}$, a trigonal distortion term $\delta(l_z^2-2/3)$, and a coupling to external magnetic field $\mu_{\mathrm{B}}\bm{B}\cdot(2\bm{S}-3\bm{l}/2)$. The magnetic interactions are of the Heisenberg type between nearest neighbors within the plane, $J_1$, and between planes, $J_4$, $\sum_{i,j}\left(J_1\bm{S}_i\cdot\bm{S}_j\right)+\sum_{i,k}\left(J_4\bm{S}_i\cdot\bm{S}_k\right)$. There is also a ring-exchange term that is responsible for the opening of the magnon gap at the zone-center, but this term does not affect the spin-orbital levels. In addition, there is a biquadratic exchange term with strength $q$, $\sum_{i,j}\left[q(S^+_i S^-_j)^2+q^*(S^-_i S^+_j)^2\right]$, that was previously proposed to explain the bandwidth of the spin-orbiton dispersion \cite{YuanPRB2020}. The full details of the calculation are given in \cite{Yufei,YufeiThesis}.

While previous measurements could not resolve the splitting between P$_1$ and P$_2$ \cite{YuanPRB2020,Elliot2021}, our higher experimental energy resolution allows us to determine the zero field splitting, which we reproduce with the following parameters: $\lambda=\SI{16.4}{meV}$, $\delta=\SI{52}{meV}$, $J_1=\SI{-0.90}{meV}$, $J_4=\SI{0.189}{meV}$, and $q=\SI{-0.15}{meV}$. By varying the values of these parameters we find that the strongest dependence of this splitting is on $J_4$, the inter-plane exchange interaction, then on the biquadratic exchange $q$, and only weakly on $J_1$. We can understand, within this model, that P$_1$ and P$_2$ are transitions between the AFM ground state and the lowest two doubly degenerate levels that result from mixing the $J_z=\pm 3/2$ single-ion states from the four Co atoms that make up the AFM unit cell. This mixing produces four doublets that split under the application of a magnetic field applied along the $c$-axis, as schematically shown in Fig. \ref{fig:1}(c). The lowest two doublets correspond to P$_1$ and P$_2$, and the highest two would correspond to the spin-orbiton measured in the infrared experiment.

Previously, a model \cite{YuanPRB2020} only considered the coupling between two Co spins and the effect of the mean field on the Co spin due to its neighbors. Our model correctly considers all of the four Co spins in the unit cell and is also able to fit the dispersion of these spin-orbitons, see \cite{YufeiThesis}. It also can reproduce the magnetic field dependence, applied along the $c$-axis shown in Fig. \ref{fig:Fitting}(b) and (c), of the P$_1$ and P$_2$ excitations. While the fit is not perfect, the overall behavior is well reproduced for the six detected excitations. A notable difference can be seen by the effective $g$-factors (linear slopes) of the spin-orbitons; they are not perfectly matched to experiment, but they do qualitatively explain the behavior of all the spin-orbitons when the mangetic field is along the $c$-axis. At the same time, our calculation does not reproduce the behavior of P$_1$ and P$_2$ when the magnetic field is perpendicular to the $c$-axis, in our Voigt geometry experiment, (Fig. \ref{fig:FieldDep}(b)). In our calculation, we found that the energy difference between P$_1$ and P$_2$ remains nearly constant with increasing magnetic field as long as the value of $J_1$ is finite.  This does not match the observation of the merging of P$_1$ and P$_2$ above 7 T.

We consider the possibility of the interaction between the spin-orbitons with phonons. From the temperature- and field-dependent data, we know that the \Eg{1} phonon is strongly coupled to the magnetism. There are signs of coupling of the zone-folded partner to \Eg{1} as well. Thus, we can conclude that the hybridization of the \Eg{1} phonon with the spin-orbitons P$_1$ and P$_2$ explains their temperature and magnetic field dependence. At the same time, the magnetic field dependence of the \Eu{} phonon also shows signs that it hybridizes with the lower branch of its neighboring spin-orbiton. From these observations, we speculate that an extended model that allows for spin-orbiton-phonon interaction would be able to address some of the differences between our current model prediction and the experimental data. 

Indeed, such a model has been developed by \citet{Chaudhary_PRB2024}. They considered the coupling between E$_g$ phonons and the single-ion $J_z=\pm 3/2$ spin-orbital levels due to the electron orbital-phonon coupling and found the Zeeman splitting of both the phonon and spin-orbitons, when the magnetic field is parallel to the $c$-axis. Qualitatively, this is consistent with the data presented in Fig. \ref{fig:Fitting}(b) and (c). However, a close comparison shows that this model is still incomplete. Their predicted splitting for \Eg{1} is too small and they do not include the zone-folded phonon near \Eg{1}. Also, the slopes (effective $g$-factors) for the Zeeman splittings of P$_1$ and P$_2$ are supposed to be the same in magnitude, but the data in Fig. \ref{fig:Fitting}(c) clearly shows a smaller magnitude of the slope for the positive one than the negative one. The former may be due to the close proximity to the lower branch of the split \Eg{1} phonon. The near degeneracy of these modes may be causing an avoided crossing at high fields which would have the effect of decreasing the slopes of these two modes. This model also does not consider the coupling to infrared active phonons, and so it cannot explain our data in Fig. \ref{fig:FTIRFieldDep}. These effects are not reproduced by the model in \cite{Chaudhary_PRB2024}, and thus further theoretical work is needed to accurately reproduce all the experimental results.

\section{Summary}

We show a detailed temperature and magnetic field dependent Raman scattering study, and the magnetic field dependence of the infrared spectra to probe the coupling between phonons and spin-orbitons in \ch{CoTiO_3}. We report three principal effects: 1) the appearance below \TN\ of 5 zone-folded phonons and their disappearance above $H_0$ perpendicular to the $c$-axis, 2) observation of the hybridization between two spin-orbitons with a nearby phonon with E$_\text{g}$ symmetry, giving rise to its Zeeman effect, and 3) the hybridization between an inversion-odd \Eu{} phonon and an additional spin-orbiton measured in the infrared spectra in magnetic field also applied along the $c$-axis. While the model that explains the magnon behavior \cite{Yufei,YufeiThesis} can explain many of the experimental results, it does not do so exactly. This highlights the need to further develop the microscopic theory of the coupling between the spin, orbital, and lattice degrees of freedom in \ch{CoTiO_3}.

\textit{Note added}. During the completion of this manuscript, we became aware of a paper \cite{Lujan2024PNASCTO} with very similar Raman data on \ch{CoTiO_3} but with a different interpretation on the nature of the excitations we label P$_1$ and P$_2$. Those authors do not consider that in the AFM state there are four Co atoms in the unit cell which significantly changes the single ion picture they considered.

\section{Acknowledgements}

Work at OSU and CSU was supported by the Center for Emergent Materials at OSU, a Materials Research Science and Engineering Center funded by NSF under grant DMR-2011876. We thank Nandini Trivedi, Yuanming Lu, and Swati Chaudhary for helpful discussions.

\section{Methods}
\textbf{Single crystal growth}. The samples used in these experiments were grown by the floating zone method as described earlier \cite{Yufei}. They are large single crystal approximately 1\,mm thick and a few mm in diameter. One of the crystals has the large surface parallel to the $a$-$b$ plane, while the other one has its large surface perpendicular to the [110] direction, as determined by Laue X-ray diffraction and polarization-dependent Raman scattering. These single crystals were previously used to study the magnon energies with THz and magneto-Raman spectroscopies \cite{Yufei}.

\textbf{Magneto-Raman spectroscopy}. The magneto-Raman experiments at NIST used two gas line lasers, with wavelengths of 633 nm (not shown here) and 488\,nm, a triple grating spectrometer, and a closed-cycle magneto cryostat with optical access. The sample is mounted inside the cryostat in a He exchange gas environment to allow temperature control down to 2\,K. The laser propagates through free space optics and is focused onto the sample at normal incidence via a non-magnetic objective. The laser spot size on the sample is in the order of $\qty{1}{\micro\meter}$ in diameter. The laser power was kept below 2.0\,mW to avoid sample damage. The superconducting magnet can apply up to 9\,T magnetic field to the sample in either the Faraday (normal to the sample surface) or Voigt (parallel to the sample surface) geometry. The incident and scattered laser polarizations can be controlled by various polarization optics. The two configurations reported in this study are VV and VH, which correspond to parallel and crossed polarizations respectively. Raman experiments with unpolarized light at 633\,nm, magnetic fields up to 22\,T parallel to the $a$-$b$ plane were performed at LNCMI in Grenoble, France with the sample kept at 5\,K. 

\textbf{Infrared spectroscopy}. IR reflection measurements in the range between 100 cm$^{-1}$ and 600 cm$^{-1}$ were performed at LNCMI using Fourier transform infrared (FTIR) spectroscopy in magnetic fields up to 16 T using the [110]-cut sample kept at 4.2\,K. Unpolarized broadband radiation was incident on the sample, and the reflected light was detected using a polyethylene polarizer oriented either along the [001] or the [1$\bar{1}$0] axis.

\bibliography{CoTiO3Raman}

\begin{thebibliography}{39}%
\makeatletter
\providecommand \@ifxundefined [1]{%
 \@ifx{#1\undefined}
}%
\providecommand \@ifnum [1]{%
 \ifnum #1\expandafter \@firstoftwo
 \else \expandafter \@secondoftwo
 \fi
}%
\providecommand \@ifx [1]{%
 \ifx #1\expandafter \@firstoftwo
 \else \expandafter \@secondoftwo
 \fi
}%
\providecommand \natexlab [1]{#1}%
\providecommand \enquote  [1]{``#1''}%
\providecommand \bibnamefont  [1]{#1}%
\providecommand \bibfnamefont [1]{#1}%
\providecommand \citenamefont [1]{#1}%
\providecommand \href@noop [0]{\@secondoftwo}%
\providecommand \href [0]{\begingroup \@sanitize@url \@href}%
\providecommand \@href[1]{\@@startlink{#1}\@@href}%
\providecommand \@@href[1]{\endgroup#1\@@endlink}%
\providecommand \@sanitize@url [0]{\catcode `\\12\catcode `\$12\catcode
  `\&12\catcode `\#12\catcode `\^12\catcode `\_12\catcode `\%12\relax}%
\providecommand \@@startlink[1]{}%
\providecommand \@@endlink[0]{}%
\providecommand \url  [0]{\begingroup\@sanitize@url \@url }%
\providecommand \@url [1]{\endgroup\@href {#1}{\urlprefix }}%
\providecommand \urlprefix  [0]{URL }%
\providecommand \Eprint [0]{\href }%
\providecommand \doibase [0]{https://doi.org/}%
\providecommand \selectlanguage [0]{\@gobble}%
\providecommand \bibinfo  [0]{\@secondoftwo}%
\providecommand \bibfield  [0]{\@secondoftwo}%
\providecommand \translation [1]{[#1]}%
\providecommand \BibitemOpen [0]{}%
\providecommand \bibitemStop [0]{}%
\providecommand \bibitemNoStop [0]{.\EOS\space}%
\providecommand \EOS [0]{\spacefactor3000\relax}%
\providecommand \BibitemShut  [1]{\csname bibitem#1\endcsname}%
\let\auto@bib@innerbib\@empty
\bibitem [{\citenamefont {Henry}\ and\ \citenamefont
  {Hopfield}(1965)}]{PhysRevLett.15.964}%
  \BibitemOpen
  \bibfield  {author} {\bibinfo {author} {\bibfnamefont {C.~H.}\ \bibnamefont
  {Henry}}\ and\ \bibinfo {author} {\bibfnamefont {J.~J.}\ \bibnamefont
  {Hopfield}},\ }\bibfield  {title} {\bibinfo {title} {Raman scattering by
  polaritons},\ }\href {https://doi.org/10.1103/PhysRevLett.15.964} {\bibfield
  {journal} {\bibinfo  {journal} {Phys. Rev. Lett.}\ }\textbf {\bibinfo
  {volume} {15}},\ \bibinfo {pages} {964} (\bibinfo {year} {1965})}\BibitemShut
  {NoStop}%
\bibitem [{\citenamefont {Rahimi-Iman}(2020)}]{PolaritonPhysics}%
  \BibitemOpen
  \bibfield  {author} {\bibinfo {author} {\bibfnamefont {A.}~\bibnamefont
  {Rahimi-Iman}},\ }\href@noop {} {\emph {\bibinfo {title} {Polariton
  Physics}}}\ (\bibinfo  {publisher} {Springer},\ \bibinfo {year}
  {2020})\BibitemShut {NoStop}%
\bibitem [{\citenamefont {Feurer}\ \emph {et~al.}(2007)\citenamefont {Feurer},
  \citenamefont {Stoyanov}, \citenamefont {Ward}, \citenamefont {Vaughan},
  \citenamefont {Statz},\ and\ \citenamefont {Nelson}}]{TerahertzPolaritonics}%
  \BibitemOpen
  \bibfield  {author} {\bibinfo {author} {\bibfnamefont {T.}~\bibnamefont
  {Feurer}}, \bibinfo {author} {\bibfnamefont {N.~S.}\ \bibnamefont
  {Stoyanov}}, \bibinfo {author} {\bibfnamefont {D.~W.}\ \bibnamefont {Ward}},
  \bibinfo {author} {\bibfnamefont {J.~C.}\ \bibnamefont {Vaughan}}, \bibinfo
  {author} {\bibfnamefont {E.~R.}\ \bibnamefont {Statz}},\ and\ \bibinfo
  {author} {\bibfnamefont {K.~A.}\ \bibnamefont {Nelson}},\ }\bibfield  {title}
  {\bibinfo {title} {Terahertz polaritonics},\ }\href
  {https://doi.org/https://doi.org/10.1146/annurev.matsci.37.052506.084327}
  {\bibfield  {journal} {\bibinfo  {journal} {Annual Review of Materials
  Research}\ }\textbf {\bibinfo {volume} {37}},\ \bibinfo {pages} {317}
  (\bibinfo {year} {2007})}\BibitemShut {NoStop}%
\bibitem [{\citenamefont {Mai}\ \emph {et~al.}(2021)\citenamefont {Mai},
  \citenamefont {Garrity}, \citenamefont {McCreary}, \citenamefont {Argo},
  \citenamefont {Simpson}, \citenamefont {Doan-Nguyen}, \citenamefont
  {Aguilar},\ and\ \citenamefont {Walker}}]{Mai2021_SciAdv}%
  \BibitemOpen
  \bibfield  {author} {\bibinfo {author} {\bibfnamefont {T.~T.}\ \bibnamefont
  {Mai}}, \bibinfo {author} {\bibfnamefont {K.~F.}\ \bibnamefont {Garrity}},
  \bibinfo {author} {\bibfnamefont {A.}~\bibnamefont {McCreary}}, \bibinfo
  {author} {\bibfnamefont {J.}~\bibnamefont {Argo}}, \bibinfo {author}
  {\bibfnamefont {J.~R.}\ \bibnamefont {Simpson}}, \bibinfo {author}
  {\bibfnamefont {V.}~\bibnamefont {Doan-Nguyen}}, \bibinfo {author}
  {\bibfnamefont {R.~V.}\ \bibnamefont {Aguilar}},\ and\ \bibinfo {author}
  {\bibfnamefont {A.~R.~H.}\ \bibnamefont {Walker}},\ }\bibfield  {title}
  {\bibinfo {title} {Magnon-phonon hybridization in 2d antiferromagnet
  mnpse<sub>3</sub>},\ }\href {https://doi.org/10.1126/sciadv.abj3106}
  {\bibfield  {journal} {\bibinfo  {journal} {Science Advances}\ }\textbf
  {\bibinfo {volume} {7}},\ \bibinfo {pages} {eabj3106} (\bibinfo {year}
  {2021})},\ \Eprint
  {https://arxiv.org/abs/https://www.science.org/doi/pdf/10.1126/sciadv.abj3106}
  {https://www.science.org/doi/pdf/10.1126/sciadv.abj3106} \BibitemShut
  {NoStop}%
\bibitem [{\citenamefont {Abragam}\ and\ \citenamefont
  {Bleaney}(1970)}]{AandBbook}%
  \BibitemOpen
  \bibfield  {author} {\bibinfo {author} {\bibfnamefont {A.}~\bibnamefont
  {Abragam}}\ and\ \bibinfo {author} {\bibfnamefont {B.}~\bibnamefont
  {Bleaney}},\ }\href@noop {} {\emph {\bibinfo {title} {Electron paramagnetic
  resonance of transition ions}}}\ (\bibinfo  {publisher} {Oxford University
  press},\ \bibinfo {year} {1970})\BibitemShut {NoStop}%
\bibitem [{\citenamefont {Sugano}(2012)}]{tanabesugano}%
  \BibitemOpen
  \bibfield  {author} {\bibinfo {author} {\bibfnamefont {S.}~\bibnamefont
  {Sugano}},\ }\href {https://books.google.com/books?id=8SbsjFx1MbwC} {\emph
  {\bibinfo {title} {Multiplets of Transition-Metal Ions in Crystals}}}\
  (\bibinfo  {publisher} {Elsevier Science},\ \bibinfo {year}
  {2012})\BibitemShut {NoStop}%
\bibitem [{\citenamefont {Khomskii}(2014)}]{Khomskii_2014}%
  \BibitemOpen
  \bibfield  {author} {\bibinfo {author} {\bibfnamefont {D.~I.}\ \bibnamefont
  {Khomskii}},\ }\href@noop {} {\emph {\bibinfo {title} {Transition Metal
  Compounds}}}\ (\bibinfo  {publisher} {Cambridge University Press},\ \bibinfo
  {year} {2014})\BibitemShut {NoStop}%
\bibitem [{\citenamefont {Yuan}\ \emph
  {et~al.}(2020{\natexlab{a}})\citenamefont {Yuan}, \citenamefont {Khait},
  \citenamefont {Shu}, \citenamefont {Chou}, \citenamefont {Stone},
  \citenamefont {Clancy}, \citenamefont {Paramekanti},\ and\ \citenamefont
  {Kim}}]{YuanPRX2020}%
  \BibitemOpen
  \bibfield  {author} {\bibinfo {author} {\bibfnamefont {B.}~\bibnamefont
  {Yuan}}, \bibinfo {author} {\bibfnamefont {I.}~\bibnamefont {Khait}},
  \bibinfo {author} {\bibfnamefont {G.-J.}\ \bibnamefont {Shu}}, \bibinfo
  {author} {\bibfnamefont {F.~C.}\ \bibnamefont {Chou}}, \bibinfo {author}
  {\bibfnamefont {M.~B.}\ \bibnamefont {Stone}}, \bibinfo {author}
  {\bibfnamefont {J.~P.}\ \bibnamefont {Clancy}}, \bibinfo {author}
  {\bibfnamefont {A.}~\bibnamefont {Paramekanti}},\ and\ \bibinfo {author}
  {\bibfnamefont {Y.-J.}\ \bibnamefont {Kim}},\ }\bibfield  {title} {\bibinfo
  {title} {Dirac magnons in a honeycomb lattice quantum $\mathit{XY}$ magnet
  ${\mathrm{cotio}}_{3}$},\ }\href {https://doi.org/10.1103/PhysRevX.10.011062}
  {\bibfield  {journal} {\bibinfo  {journal} {Phys. Rev. X}\ }\textbf {\bibinfo
  {volume} {10}},\ \bibinfo {pages} {011062} (\bibinfo {year}
  {2020}{\natexlab{a}})}\BibitemShut {NoStop}%
\bibitem [{\citenamefont {Elliot}\ \emph {et~al.}(2021)\citenamefont {Elliot},
  \citenamefont {McClarty}, \citenamefont {Prabhakaran}, \citenamefont
  {Johnson}, \citenamefont {Walker}, \citenamefont {Manuel},\ and\
  \citenamefont {Coldea}}]{Elliot2021}%
  \BibitemOpen
  \bibfield  {author} {\bibinfo {author} {\bibfnamefont {M.}~\bibnamefont
  {Elliot}}, \bibinfo {author} {\bibfnamefont {P.~A.}\ \bibnamefont
  {McClarty}}, \bibinfo {author} {\bibfnamefont {D.}~\bibnamefont
  {Prabhakaran}}, \bibinfo {author} {\bibfnamefont {R.~D.}\ \bibnamefont
  {Johnson}}, \bibinfo {author} {\bibfnamefont {H.~C.}\ \bibnamefont {Walker}},
  \bibinfo {author} {\bibfnamefont {P.}~\bibnamefont {Manuel}},\ and\ \bibinfo
  {author} {\bibfnamefont {R.}~\bibnamefont {Coldea}},\ }\bibfield  {title}
  {\bibinfo {title} {Order-by-disorder from bond-dependent exchange and
  intensity signature of nodal quasiparticles in a honeycomb cobaltate},\
  }\href {https://doi.org/10.1038/s41467-021-23851-0} {\bibfield  {journal}
  {\bibinfo  {journal} {Nature Communications}\ }\textbf {\bibinfo {volume}
  {12}},\ \bibinfo {pages} {3936} (\bibinfo {year} {2021})}\BibitemShut
  {NoStop}%
\bibitem [{\citenamefont {Li}\ \emph {et~al.}(2024)\citenamefont {Li},
  \citenamefont {Mai}, \citenamefont {Karaki}, \citenamefont {Jasper},
  \citenamefont {Garrity}, \citenamefont {Lyon}, \citenamefont {Shaw},
  \citenamefont {DeLazzer}, \citenamefont {Biacchi}, \citenamefont {Dally},
  \citenamefont {Heligman}, \citenamefont {Gdanski}, \citenamefont {Adel},
  \citenamefont {Mu\~noz}, \citenamefont {Giovannone}, \citenamefont {Pawbake},
  \citenamefont {Faugeras}, \citenamefont {Simpson}, \citenamefont {Ross},
  \citenamefont {Trivedi}, \citenamefont {Lu}, \citenamefont {Hight~Walker},\
  and\ \citenamefont {Vald\'es~Aguilar}}]{Yufei}%
  \BibitemOpen
  \bibfield  {author} {\bibinfo {author} {\bibfnamefont {Y.}~\bibnamefont
  {Li}}, \bibinfo {author} {\bibfnamefont {T.~T.}\ \bibnamefont {Mai}},
  \bibinfo {author} {\bibfnamefont {M.}~\bibnamefont {Karaki}}, \bibinfo
  {author} {\bibfnamefont {E.~V.}\ \bibnamefont {Jasper}}, \bibinfo {author}
  {\bibfnamefont {K.~F.}\ \bibnamefont {Garrity}}, \bibinfo {author}
  {\bibfnamefont {C.}~\bibnamefont {Lyon}}, \bibinfo {author} {\bibfnamefont
  {D.}~\bibnamefont {Shaw}}, \bibinfo {author} {\bibfnamefont {T.}~\bibnamefont
  {DeLazzer}}, \bibinfo {author} {\bibfnamefont {A.~J.}\ \bibnamefont
  {Biacchi}}, \bibinfo {author} {\bibfnamefont {R.~L.}\ \bibnamefont {Dally}},
  \bibinfo {author} {\bibfnamefont {D.~M.}\ \bibnamefont {Heligman}}, \bibinfo
  {author} {\bibfnamefont {J.}~\bibnamefont {Gdanski}}, \bibinfo {author}
  {\bibfnamefont {T.}~\bibnamefont {Adel}}, \bibinfo {author} {\bibfnamefont
  {M.~F.}\ \bibnamefont {Mu\~noz}}, \bibinfo {author} {\bibfnamefont
  {A.}~\bibnamefont {Giovannone}}, \bibinfo {author} {\bibfnamefont
  {A.}~\bibnamefont {Pawbake}}, \bibinfo {author} {\bibfnamefont
  {C.}~\bibnamefont {Faugeras}}, \bibinfo {author} {\bibfnamefont {J.~R.}\
  \bibnamefont {Simpson}}, \bibinfo {author} {\bibfnamefont {K.}~\bibnamefont
  {Ross}}, \bibinfo {author} {\bibfnamefont {N.}~\bibnamefont {Trivedi}},
  \bibinfo {author} {\bibfnamefont {Y.~M.}\ \bibnamefont {Lu}}, \bibinfo
  {author} {\bibfnamefont {A.~R.}\ \bibnamefont {Hight~Walker}},\ and\ \bibinfo
  {author} {\bibfnamefont {R.}~\bibnamefont {Vald\'es~Aguilar}},\ }\bibfield
  {title} {\bibinfo {title} {Ring-exchange interaction effects on magnons in
  dirac magnet \ch{CoTiO_3}},\ }\href@noop {} {\bibfield  {journal} {\bibinfo
  {journal} {Phys. Rev. B}\ }\textbf {\bibinfo {volume} {109}},\ \bibinfo
  {pages} {184436} (\bibinfo {year} {2024})}\BibitemShut {NoStop}%
\bibitem [{\citenamefont {Yuan}\ \emph
  {et~al.}(2020{\natexlab{b}})\citenamefont {Yuan}, \citenamefont {Stone},
  \citenamefont {Shu}, \citenamefont {Chou}, \citenamefont {Rao}, \citenamefont
  {Clancy},\ and\ \citenamefont {Kim}}]{YuanPRB2020}%
  \BibitemOpen
  \bibfield  {author} {\bibinfo {author} {\bibfnamefont {B.}~\bibnamefont
  {Yuan}}, \bibinfo {author} {\bibfnamefont {M.~B.}\ \bibnamefont {Stone}},
  \bibinfo {author} {\bibfnamefont {G.-J.}\ \bibnamefont {Shu}}, \bibinfo
  {author} {\bibfnamefont {F.~C.}\ \bibnamefont {Chou}}, \bibinfo {author}
  {\bibfnamefont {X.}~\bibnamefont {Rao}}, \bibinfo {author} {\bibfnamefont
  {J.~P.}\ \bibnamefont {Clancy}},\ and\ \bibinfo {author} {\bibfnamefont
  {Y.-J.}\ \bibnamefont {Kim}},\ }\bibfield  {title} {\bibinfo {title}
  {Spin-orbit exciton in a honeycomb lattice magnet ${\mathrm{cotio}}_{3}$:
  Revealing a link between magnetism in $d$- and $f$-electron systems},\ }\href
  {https://doi.org/10.1103/PhysRevB.102.134404} {\bibfield  {journal} {\bibinfo
   {journal} {Phys. Rev. B}\ }\textbf {\bibinfo {volume} {102}},\ \bibinfo
  {pages} {134404} (\bibinfo {year} {2020}{\natexlab{b}})}\BibitemShut
  {NoStop}%
\bibitem [{\citenamefont {Mittelst\"adt}\ \emph {et~al.}(2015)\citenamefont
  {Mittelst\"adt}, \citenamefont {Schmidt}, \citenamefont {Wang}, \citenamefont
  {Mayr}, \citenamefont {Tsurkan}, \citenamefont {Lunkenheimer}, \citenamefont
  {Ish}, \citenamefont {Balents}, \citenamefont {Deisenhofer},\ and\
  \citenamefont {Loidl}}]{spinOrbiton_2015}%
  \BibitemOpen
  \bibfield  {author} {\bibinfo {author} {\bibfnamefont {L.}~\bibnamefont
  {Mittelst\"adt}}, \bibinfo {author} {\bibfnamefont {M.}~\bibnamefont
  {Schmidt}}, \bibinfo {author} {\bibfnamefont {Z.}~\bibnamefont {Wang}},
  \bibinfo {author} {\bibfnamefont {F.}~\bibnamefont {Mayr}}, \bibinfo {author}
  {\bibfnamefont {V.}~\bibnamefont {Tsurkan}}, \bibinfo {author} {\bibfnamefont
  {P.}~\bibnamefont {Lunkenheimer}}, \bibinfo {author} {\bibfnamefont
  {D.}~\bibnamefont {Ish}}, \bibinfo {author} {\bibfnamefont {L.}~\bibnamefont
  {Balents}}, \bibinfo {author} {\bibfnamefont {J.}~\bibnamefont
  {Deisenhofer}},\ and\ \bibinfo {author} {\bibfnamefont {A.}~\bibnamefont
  {Loidl}},\ }\bibfield  {title} {\bibinfo {title} {Spin-orbiton and quantum
  criticality in ${\mathrm{fesc}}_{2}{\mathrm{s}}_{4}$},\ }\href
  {https://doi.org/10.1103/PhysRevB.91.125112} {\bibfield  {journal} {\bibinfo
  {journal} {Phys. Rev. B}\ }\textbf {\bibinfo {volume} {91}},\ \bibinfo
  {pages} {125112} (\bibinfo {year} {2015})}\BibitemShut {NoStop}%
\bibitem [{\citenamefont {Juraschek}\ and\ \citenamefont
  {Spaldin}(2019)}]{Juraschek_2019}%
  \BibitemOpen
  \bibfield  {author} {\bibinfo {author} {\bibfnamefont {D.~M.}\ \bibnamefont
  {Juraschek}}\ and\ \bibinfo {author} {\bibfnamefont {N.~A.}\ \bibnamefont
  {Spaldin}},\ }\bibfield  {title} {\bibinfo {title} {Orbital magnetic moments
  of phonons},\ }\href {https://doi.org/10.1103/PhysRevMaterials.3.064405}
  {\bibfield  {journal} {\bibinfo  {journal} {Phys. Rev. Mater.}\ }\textbf
  {\bibinfo {volume} {3}},\ \bibinfo {pages} {064405} (\bibinfo {year}
  {2019})}\BibitemShut {NoStop}%
\bibitem [{\citenamefont {Juraschek}\ \emph {et~al.}(2017)\citenamefont
  {Juraschek}, \citenamefont {Fechner}, \citenamefont {Balatsky},\ and\
  \citenamefont {Spaldin}}]{Juraschek2017}%
  \BibitemOpen
  \bibfield  {author} {\bibinfo {author} {\bibfnamefont {D.~M.}\ \bibnamefont
  {Juraschek}}, \bibinfo {author} {\bibfnamefont {M.}~\bibnamefont {Fechner}},
  \bibinfo {author} {\bibfnamefont {A.~V.}\ \bibnamefont {Balatsky}},\ and\
  \bibinfo {author} {\bibfnamefont {N.~A.}\ \bibnamefont {Spaldin}},\
  }\bibfield  {title} {\bibinfo {title} {Dynamical multiferroicity},\ }\href
  {https://doi.org/10.1103/PhysRevMaterials.1.014401} {\bibfield  {journal}
  {\bibinfo  {journal} {Phys. Rev. Mater.}\ }\textbf {\bibinfo {volume} {1}},\
  \bibinfo {pages} {014401} (\bibinfo {year} {2017})}\BibitemShut {NoStop}%
\bibitem [{\citenamefont {Juraschek}\ \emph {et~al.}(2020)\citenamefont
  {Juraschek}, \citenamefont {Narang},\ and\ \citenamefont
  {Spaldin}}]{Juraschek2020}%
  \BibitemOpen
  \bibfield  {author} {\bibinfo {author} {\bibfnamefont {D.~M.}\ \bibnamefont
  {Juraschek}}, \bibinfo {author} {\bibfnamefont {P.}~\bibnamefont {Narang}},\
  and\ \bibinfo {author} {\bibfnamefont {N.~A.}\ \bibnamefont {Spaldin}},\
  }\bibfield  {title} {\bibinfo {title} {Phono-magnetic analogs to
  opto-magnetic effects},\ }\href
  {https://doi.org/10.1103/PhysRevResearch.2.043035} {\bibfield  {journal}
  {\bibinfo  {journal} {Phys. Rev. Res.}\ }\textbf {\bibinfo {volume} {2}},\
  \bibinfo {pages} {043035} (\bibinfo {year} {2020})}\BibitemShut {NoStop}%
\bibitem [{\citenamefont {Schaack}(1975)}]{schaack_magnetic-field_1975}%
  \BibitemOpen
  \bibfield  {author} {\bibinfo {author} {\bibfnamefont {G.}~\bibnamefont
  {Schaack}},\ }\bibfield  {title} {\bibinfo {title} {Magnetic-field dependent
  phonon states in paramagnetic {CeF3}},\ }\href
  {https://doi.org/https://doi.org/10.1016/0038-1098(75)90488-3} {\bibfield
  {journal} {\bibinfo  {journal} {Solid State Communications}\ }\textbf
  {\bibinfo {volume} {17}},\ \bibinfo {pages} {505} (\bibinfo {year}
  {1975})}\BibitemShut {NoStop}%
\bibitem [{\citenamefont {Schaack}(1976)}]{Schaack_1976}%
  \BibitemOpen
  \bibfield  {author} {\bibinfo {author} {\bibfnamefont {G.}~\bibnamefont
  {Schaack}},\ }\bibfield  {title} {\bibinfo {title} {Observation of circularly
  polarized phonon states in an external magnetic field},\ }\href
  {https://doi.org/10.1088/0022-3719/9/11/009} {\bibfield  {journal} {\bibinfo
  {journal} {Journal of Physics C: Solid State Physics}\ }\textbf {\bibinfo
  {volume} {9}},\ \bibinfo {pages} {L297} (\bibinfo {year} {1976})}\BibitemShut
  {NoStop}%
\bibitem [{\citenamefont {Schaack}(1977)}]{Schaack1977}%
  \BibitemOpen
  \bibfield  {author} {\bibinfo {author} {\bibfnamefont {G.}~\bibnamefont
  {Schaack}},\ }\bibfield  {title} {\bibinfo {title} {Magnetic field dependent
  splitting of doubly degenerate phonon states in anhydrous
  cerium-trichloride},\ }\href {https://doi.org/10.1007/BF01313371} {\bibfield
  {journal} {\bibinfo  {journal} {Zeitschrift f{\"u}r Physik B Condensed
  Matter}\ }\textbf {\bibinfo {volume} {26}},\ \bibinfo {pages} {49} (\bibinfo
  {year} {1977})}\BibitemShut {NoStop}%
\bibitem [{\citenamefont {Janssen}(2013)}]{ITCD1}%
  \BibitemOpen
  \bibfield  {author} {\bibinfo {author} {\bibfnamefont {T.}~\bibnamefont
  {Janssen}},\ }\href {https://it.iucr.org/Db/ch1o2v0001/sec1o2o6/} {\emph
  {\bibinfo {title} {International Tables for Crystallography}}},\
  Vol.~\bibinfo {volume} {D}\ (\bibinfo {year} {2013})\ Chap.\ \bibinfo
  {chapter} {1.2}, pp.\ \bibinfo {pages} {56--62}\BibitemShut {NoStop}%
\bibitem [{\citenamefont {Bradley}\ and\ \citenamefont
  {Cracknell}(1972)}]{charactertable}%
  \BibitemOpen
  \bibfield  {author} {\bibinfo {author} {\bibfnamefont {C.}~\bibnamefont
  {Bradley}}\ and\ \bibinfo {author} {\bibfnamefont {A.}~\bibnamefont
  {Cracknell}},\ }\href@noop {} {\emph {\bibinfo {title} {The Mathematical
  Theory of Symmetry in Solids}}}\ (\bibinfo  {publisher} {Clarendon Press -
  Oxford},\ \bibinfo {year} {1972})\BibitemShut {NoStop}%
\bibitem [{\citenamefont {Gregora}(2013)}]{ITCD2}%
  \BibitemOpen
  \bibfield  {author} {\bibinfo {author} {\bibfnamefont {I.}~\bibnamefont
  {Gregora}},\ }\href {https://it.iucr.org/Db/ch2o3v0001/} {\emph {\bibinfo
  {title} {International Tables for Crystallography}}},\ Vol.~\bibinfo {volume}
  {D}\ (\bibinfo {year} {2013})\ Chap.\ \bibinfo {chapter} {2.3}, pp.\ \bibinfo
  {pages} {334--348}\BibitemShut {NoStop}%
\bibitem [{\citenamefont {Ovander}(1960)}]{Ovander1960}%
  \BibitemOpen
  \bibfield  {author} {\bibinfo {author} {\bibfnamefont {L.}~\bibnamefont
  {Ovander}},\ }\bibfield  {title} {\bibinfo {title} {The form of the raman
  tensor},\ }\href@noop {} {\bibfield  {journal} {\bibinfo  {journal} {Optics
  and Spectroscopy}\ }\textbf {\bibinfo {volume} {9}},\ \bibinfo {pages} {302}
  (\bibinfo {year} {1960})}\BibitemShut {NoStop}%
\bibitem [{\citenamefont {Anastassakis}\ \emph {et~al.}(1972)\citenamefont
  {Anastassakis}, \citenamefont {Burstein}, \citenamefont {Maradudin},\ and\
  \citenamefont {Minnick}}]{ANASTASSAKIS1972}%
  \BibitemOpen
  \bibfield  {author} {\bibinfo {author} {\bibfnamefont {E.}~\bibnamefont
  {Anastassakis}}, \bibinfo {author} {\bibfnamefont {E.}~\bibnamefont
  {Burstein}}, \bibinfo {author} {\bibfnamefont {A.}~\bibnamefont
  {Maradudin}},\ and\ \bibinfo {author} {\bibfnamefont {R.}~\bibnamefont
  {Minnick}},\ }\bibfield  {title} {\bibinfo {title} {Morphic effects — iii.
  effects of an external magnetic field on the long wavelength optical
  phonons},\ }\href
  {https://doi.org/https://doi.org/10.1016/0022-3697(72)90034-0} {\bibfield
  {journal} {\bibinfo  {journal} {Journal of Physics and Chemistry of Solids}\
  }\textbf {\bibinfo {volume} {33}},\ \bibinfo {pages} {519} (\bibinfo {year}
  {1972})}\BibitemShut {NoStop}%
\bibitem [{\citenamefont {Newnham}\ \emph {et~al.}(1964)\citenamefont
  {Newnham}, \citenamefont {Fang},\ and\ \citenamefont
  {Santoro}}]{Newnham:a04103}%
  \BibitemOpen
  \bibfield  {author} {\bibinfo {author} {\bibfnamefont {R.~E.}\ \bibnamefont
  {Newnham}}, \bibinfo {author} {\bibfnamefont {J.~H.}\ \bibnamefont {Fang}},\
  and\ \bibinfo {author} {\bibfnamefont {R.~P.}\ \bibnamefont {Santoro}},\
  }\bibfield  {title} {\bibinfo {title} {{Crystal structure and magnetic
  properties of CoTiO${\sb 3}$}},\ }\href
  {https://doi.org/10.1107/S0365110X64000615} {\bibfield  {journal} {\bibinfo
  {journal} {Acta Crystallographica}\ }\textbf {\bibinfo {volume} {17}},\
  \bibinfo {pages} {240} (\bibinfo {year} {1964})}\BibitemShut {NoStop}%
\bibitem [{\citenamefont {Liu}\ and\ \citenamefont
  {Khaliullin}(2018)}]{liuKhaliullin}%
  \BibitemOpen
  \bibfield  {author} {\bibinfo {author} {\bibfnamefont {H.}~\bibnamefont
  {Liu}}\ and\ \bibinfo {author} {\bibfnamefont {G.}~\bibnamefont
  {Khaliullin}},\ }\bibfield  {title} {\bibinfo {title} {Pseudospin exchange
  interactions in ${d}^{7}$ cobalt compounds: Possible realization of the
  kitaev model},\ }\href {https://doi.org/10.1103/PhysRevB.97.014407}
  {\bibfield  {journal} {\bibinfo  {journal} {Phys. Rev. B}\ }\textbf {\bibinfo
  {volume} {97}},\ \bibinfo {pages} {014407} (\bibinfo {year}
  {2018})}\BibitemShut {NoStop}%
\bibitem [{\citenamefont {Chaudhary}\ \emph {et~al.}(2024)\citenamefont
  {Chaudhary}, \citenamefont {Juraschek}, \citenamefont {Rodriguez-Vega},\ and\
  \citenamefont {Fiete}}]{Chaudhary_PRB2024}%
  \BibitemOpen
  \bibfield  {author} {\bibinfo {author} {\bibfnamefont {S.}~\bibnamefont
  {Chaudhary}}, \bibinfo {author} {\bibfnamefont {D.~M.}\ \bibnamefont
  {Juraschek}}, \bibinfo {author} {\bibfnamefont {M.}~\bibnamefont
  {Rodriguez-Vega}},\ and\ \bibinfo {author} {\bibfnamefont {G.~A.}\
  \bibnamefont {Fiete}},\ }\bibfield  {title} {\bibinfo {title} {Giant
  effective magnetic moments of chiral phonons from orbit-lattice coupling},\
  }\href {https://doi.org/10.1103/PhysRevB.110.094401} {\bibfield  {journal}
  {\bibinfo  {journal} {Phys. Rev. B}\ }\textbf {\bibinfo {volume} {110}},\
  \bibinfo {pages} {094401} (\bibinfo {year} {2024})}\BibitemShut {NoStop}%
\bibitem [{\citenamefont {Dubrovin}\ \emph {et~al.}(2021)\citenamefont
  {Dubrovin}, \citenamefont {Siverin}, \citenamefont {Prosnikov}, \citenamefont
  {Chernyshev}, \citenamefont {Novikova}, \citenamefont {Christianen},
  \citenamefont {Balbashov},\ and\ \citenamefont {Pisarev}}]{DUBROVIN2021}%
  \BibitemOpen
  \bibfield  {author} {\bibinfo {author} {\bibfnamefont {R.}~\bibnamefont
  {Dubrovin}}, \bibinfo {author} {\bibfnamefont {N.}~\bibnamefont {Siverin}},
  \bibinfo {author} {\bibfnamefont {M.}~\bibnamefont {Prosnikov}}, \bibinfo
  {author} {\bibfnamefont {V.}~\bibnamefont {Chernyshev}}, \bibinfo {author}
  {\bibfnamefont {N.}~\bibnamefont {Novikova}}, \bibinfo {author}
  {\bibfnamefont {P.}~\bibnamefont {Christianen}}, \bibinfo {author}
  {\bibfnamefont {A.}~\bibnamefont {Balbashov}},\ and\ \bibinfo {author}
  {\bibfnamefont {R.}~\bibnamefont {Pisarev}},\ }\bibfield  {title} {\bibinfo
  {title} {Lattice dynamics and spontaneous magnetodielectric effect in
  ilmenite cotio3},\ }\href
  {https://doi.org/https://doi.org/10.1016/j.jallcom.2020.157633} {\bibfield
  {journal} {\bibinfo  {journal} {Journal of Alloys and Compounds}\ }\textbf
  {\bibinfo {volume} {858}},\ \bibinfo {pages} {157633} (\bibinfo {year}
  {2021})}\BibitemShut {NoStop}%
\bibitem [{\citenamefont {Hoffmann}\ \emph {et~al.}(2021)\citenamefont
  {Hoffmann}, \citenamefont {Dey}, \citenamefont {Werner}, \citenamefont {Bag},
  \citenamefont {Kaiser}, \citenamefont {Wadepohl}, \citenamefont {Skourski},
  \citenamefont {Abdel-Hafiez}, \citenamefont {Singh},\ and\ \citenamefont
  {Klingeler}}]{permeability2021}%
  \BibitemOpen
  \bibfield  {author} {\bibinfo {author} {\bibfnamefont {M.}~\bibnamefont
  {Hoffmann}}, \bibinfo {author} {\bibfnamefont {K.}~\bibnamefont {Dey}},
  \bibinfo {author} {\bibfnamefont {J.}~\bibnamefont {Werner}}, \bibinfo
  {author} {\bibfnamefont {R.}~\bibnamefont {Bag}}, \bibinfo {author}
  {\bibfnamefont {J.}~\bibnamefont {Kaiser}}, \bibinfo {author} {\bibfnamefont
  {H.}~\bibnamefont {Wadepohl}}, \bibinfo {author} {\bibfnamefont
  {Y.}~\bibnamefont {Skourski}}, \bibinfo {author} {\bibfnamefont
  {M.}~\bibnamefont {Abdel-Hafiez}}, \bibinfo {author} {\bibfnamefont
  {S.}~\bibnamefont {Singh}},\ and\ \bibinfo {author} {\bibfnamefont
  {R.}~\bibnamefont {Klingeler}},\ }\bibfield  {title} {\bibinfo {title}
  {Magnetic phase diagram, magnetoelastic coupling, and gr\"uneisen scaling in
  {CoTiO}$_{3}$},\ }\href {https://doi.org/10.1103/PhysRevB.104.014429}
  {\bibfield  {journal} {\bibinfo  {journal} {Phys. Rev. B}\ }\textbf {\bibinfo
  {volume} {104}},\ \bibinfo {pages} {014429} (\bibinfo {year}
  {2021})}\BibitemShut {NoStop}%
\bibitem [{\citenamefont {Hohenberg}\ and\ \citenamefont {Kohn}(1964)}]{hk}%
  \BibitemOpen
  \bibfield  {author} {\bibinfo {author} {\bibfnamefont {P.}~\bibnamefont
  {Hohenberg}}\ and\ \bibinfo {author} {\bibfnamefont {W.}~\bibnamefont
  {Kohn}},\ }\href@noop {} {\bibfield  {journal} {\bibinfo  {journal} {Phys.\
  Rev.}\ }\textbf {\bibinfo {volume} {136}},\ \bibinfo {pages} {B864} (\bibinfo
  {year} {1964})}\BibitemShut {NoStop}%
\bibitem [{\citenamefont {Kohn}\ and\ \citenamefont {Sham}(1965)}]{ks}%
  \BibitemOpen
  \bibfield  {author} {\bibinfo {author} {\bibfnamefont {W.}~\bibnamefont
  {Kohn}}\ and\ \bibinfo {author} {\bibfnamefont {L.}~\bibnamefont {Sham}},\
  }\href@noop {} {\bibfield  {journal} {\bibinfo  {journal} {Phys.\ Rev.}\
  }\textbf {\bibinfo {volume} {140}},\ \bibinfo {pages} {A1133} (\bibinfo
  {year} {1965})}\BibitemShut {NoStop}%
\bibitem [{\citenamefont {Togo}(2023{\natexlab{a}})}]{phonopyNew}%
  \BibitemOpen
  \bibfield  {author} {\bibinfo {author} {\bibfnamefont {A.}~\bibnamefont
  {Togo}},\ }\bibfield  {title} {\bibinfo {title} {First-principles phonon
  calculations with phonopy and phono3py},\ }\href
  {https://doi.org/10.7566/JPSJ.92.012001} {\bibfield  {journal} {\bibinfo
  {journal} {Journal of the Physical Society of Japan}\ }\textbf {\bibinfo
  {volume} {92}},\ \bibinfo {pages} {012001} (\bibinfo {year}
  {2023}{\natexlab{a}})},\ \Eprint
  {https://arxiv.org/abs/https://doi.org/10.7566/JPSJ.92.012001}
  {https://doi.org/10.7566/JPSJ.92.012001} \BibitemShut {NoStop}%
\bibitem [{\citenamefont {{\textit et. al.}}(2020)}]{qe}%
  \BibitemOpen
  \bibfield  {author} {\bibinfo {author} {\bibfnamefont {G.~P.}\ \bibnamefont
  {{\textit et. al.}}},\ }\bibfield  {title} {\bibinfo {title} {Quantum
  espresso toward the exascale},\ }\href {https://doi.org/10.1063/5.0005082}
  {\bibfield  {journal} {\bibinfo  {journal} {The Journal of Chemical Physics}\
  }\textbf {\bibinfo {volume} {152}},\ \bibinfo {pages} {154105} (\bibinfo
  {year} {2020})}\BibitemShut {NoStop}%
\bibitem [{\citenamefont {Garrity}\ \emph {et~al.}(2014)\citenamefont
  {Garrity}, \citenamefont {Bennett}, \citenamefont {Rabe},\ and\ \citenamefont
  {Vanderbilt}}]{gbrv}%
  \BibitemOpen
  \bibfield  {author} {\bibinfo {author} {\bibfnamefont {K.~F.}\ \bibnamefont
  {Garrity}}, \bibinfo {author} {\bibfnamefont {J.~W.}\ \bibnamefont
  {Bennett}}, \bibinfo {author} {\bibfnamefont {K.~M.}\ \bibnamefont {Rabe}},\
  and\ \bibinfo {author} {\bibfnamefont {D.}~\bibnamefont {Vanderbilt}},\
  }\href@noop {} {\bibfield  {journal} {\bibinfo  {journal} {Comput. Mater.
  Sci}\ }\textbf {\bibinfo {volume} {81}},\ \bibinfo {pages} {446} (\bibinfo
  {year} {2014})}\BibitemShut {NoStop}%
\bibitem [{\citenamefont {Perdew}\ \emph {et~al.}(2008)\citenamefont {Perdew},
  \citenamefont {Ruzsinszky}, \citenamefont {Csonka}, \citenamefont {Vydrov},\
  and\ \citenamefont {Scus}}]{pbesol}%
  \BibitemOpen
  \bibfield  {author} {\bibinfo {author} {\bibfnamefont {J.~P.}\ \bibnamefont
  {Perdew}}, \bibinfo {author} {\bibfnamefont {A.}~\bibnamefont {Ruzsinszky}},
  \bibinfo {author} {\bibfnamefont {G.~I.}\ \bibnamefont {Csonka}}, \bibinfo
  {author} {\bibfnamefont {O.~A.}\ \bibnamefont {Vydrov}},\ and\ \bibinfo
  {author} {\bibfnamefont {G.~E.}\ \bibnamefont {Scus}},\ }\bibfield  {title}
  {\bibinfo {title} {Restoring the density-gradient expansion for exchange in
  solids and surfaces},\ }\href
  {https://doi.org/10.1103/PhysRevLett.100.136406} {\bibfield  {journal}
  {\bibinfo  {journal} {Phys. Rev. Lett.}\ }\textbf {\bibinfo {volume} {100}},\
  \bibinfo {pages} {136406} (\bibinfo {year} {2008})}\BibitemShut {NoStop}%
\bibitem [{\citenamefont {Cococcioni}(2012)}]{DFTU}%
  \BibitemOpen
  \bibfield  {author} {\bibinfo {author} {\bibfnamefont {M.}~\bibnamefont
  {Cococcioni}},\ }\bibfield  {title} {\bibinfo {title} {The lda+ u approach: a
  simple hubbard correction for correlated ground states},\ }\href@noop {}
  {\bibfield  {journal} {\bibinfo  {journal} {Correlated Electrons: From Models
  to Materials Modeling and Simulation; Verlag des Forschungszentrum
  J{\"u}lich: J{\"u}lich, Germany}\ } (\bibinfo {year} {2012})}\BibitemShut
  {NoStop}%
\bibitem [{\citenamefont {Togo}\ \emph {et~al.}(2023)\citenamefont {Togo},
  \citenamefont {Chaput}, \citenamefont {Tadano},\ and\ \citenamefont
  {Tanaka}}]{phonopy-phono3py-JPCM}%
  \BibitemOpen
  \bibfield  {author} {\bibinfo {author} {\bibfnamefont {A.}~\bibnamefont
  {Togo}}, \bibinfo {author} {\bibfnamefont {L.}~\bibnamefont {Chaput}},
  \bibinfo {author} {\bibfnamefont {T.}~\bibnamefont {Tadano}},\ and\ \bibinfo
  {author} {\bibfnamefont {I.}~\bibnamefont {Tanaka}},\ }\bibfield  {title}
  {\bibinfo {title} {Implementation strategies in phonopy and phono3py},\
  }\href {https://doi.org/10.1088/1361-648X/acd831} {\bibfield  {journal}
  {\bibinfo  {journal} {J. Phys. Condens. Matter}\ }\textbf {\bibinfo {volume}
  {35}},\ \bibinfo {pages} {353001} (\bibinfo {year} {2023})}\BibitemShut
  {NoStop}%
\bibitem [{\citenamefont {Togo}(2023{\natexlab{b}})}]{phonopy-phono3py-JPSJ}%
  \BibitemOpen
  \bibfield  {author} {\bibinfo {author} {\bibfnamefont {A.}~\bibnamefont
  {Togo}},\ }\bibfield  {title} {\bibinfo {title} {First-principles phonon
  calculations with phonopy and phono3py},\ }\href
  {https://doi.org/10.7566/JPSJ.92.012001} {\bibfield  {journal} {\bibinfo
  {journal} {J. Phys. Soc. Jpn.}\ }\textbf {\bibinfo {volume} {92}},\ \bibinfo
  {pages} {012001} (\bibinfo {year} {2023}{\natexlab{b}})}\BibitemShut
  {NoStop}%
\bibitem [{\citenamefont {Li}(2024)}]{YufeiThesis}%
  \BibitemOpen
  \bibfield  {author} {\bibinfo {author} {\bibfnamefont {Y.}~\bibnamefont
  {Li}},\ }\emph {\bibinfo {title} {Terahertz and Raman spectroscopy studies on
  quantum magnets}},\ \href@noop {} {Ph.D. thesis},\ \bibinfo  {school} {The
  Ohio State University} (\bibinfo {year} {2024})\BibitemShut {NoStop}%
\bibitem [{\citenamefont {Lujan}\ \emph {et~al.}(2024)\citenamefont {Lujan},
  \citenamefont {Choe}, \citenamefont {Chaudhary}, \citenamefont {Ye},
  \citenamefont {Nnokwe}, \citenamefont {Rodriguez-Vega}, \citenamefont {He},
  \citenamefont {Gao}, \citenamefont {Nunley}, \citenamefont {Baldini},
  \citenamefont {Zhou}, \citenamefont {Fiete}, \citenamefont {He},\ and\
  \citenamefont {Li}}]{Lujan2024PNASCTO}%
  \BibitemOpen
  \bibfield  {author} {\bibinfo {author} {\bibfnamefont {D.}~\bibnamefont
  {Lujan}}, \bibinfo {author} {\bibfnamefont {J.}~\bibnamefont {Choe}},
  \bibinfo {author} {\bibfnamefont {S.}~\bibnamefont {Chaudhary}}, \bibinfo
  {author} {\bibfnamefont {G.}~\bibnamefont {Ye}}, \bibinfo {author}
  {\bibfnamefont {C.}~\bibnamefont {Nnokwe}}, \bibinfo {author} {\bibfnamefont
  {M.}~\bibnamefont {Rodriguez-Vega}}, \bibinfo {author} {\bibfnamefont
  {J.}~\bibnamefont {He}}, \bibinfo {author} {\bibfnamefont {F.~Y.}\
  \bibnamefont {Gao}}, \bibinfo {author} {\bibfnamefont {T.~N.}\ \bibnamefont
  {Nunley}}, \bibinfo {author} {\bibfnamefont {E.}~\bibnamefont {Baldini}},
  \bibinfo {author} {\bibfnamefont {J.}~\bibnamefont {Zhou}}, \bibinfo {author}
  {\bibfnamefont {G.~A.}\ \bibnamefont {Fiete}}, \bibinfo {author}
  {\bibfnamefont {R.}~\bibnamefont {He}},\ and\ \bibinfo {author}
  {\bibfnamefont {X.}~\bibnamefont {Li}},\ }\bibfield  {title} {\bibinfo
  {title} {Spin–orbit exciton–induced phonon chirality in a quantum
  magnet},\ }\href {https://doi.org/10.1073/pnas.2304360121} {\bibfield
  {journal} {\bibinfo  {journal} {Proceedings of the National Academy of
  Sciences}\ }\textbf {\bibinfo {volume} {121}},\ \bibinfo {pages}
  {e2304360121} (\bibinfo {year} {2024})},\ \Eprint
  {https://arxiv.org/abs/https://www.pnas.org/doi/pdf/10.1073/pnas.2304360121}
  {https://www.pnas.org/doi/pdf/10.1073/pnas.2304360121} \BibitemShut {NoStop}%
\end{thebibliography}%

\end{document}